\documentclass[11pt,a4paper]{article} 
\pdfoutput=1
\usepackage{multirow,slashed,jheppub, hyperref,mdwlist}
\usepackage[utf8]{inputenc}
\usepackage{color}
\usepackage{soul}
\usepackage{eufrak}
\pdfoutput=1
\usepackage{threeparttable}
\usepackage{ulem}
\usepackage{float}

\newcommand{\be}{\begin{equation}}
\newcommand{\ee}{\end{equation}}
\newcommand{\bea}{\begin{eqnarray}}
\newcommand{\eea}{\end{eqnarray}}
\newcommand{\hc}{\text{h.c.}}
\newcommand{\Br}{\text{Br}}
\newcommand{\nn}{\nonumber}
\def\bsp#1\esp{\begin{split}#1\end{split}}

\title{Exotic decays of top partners with charge 5/3: bounds and opportunities}

\author[1]{Ke-Pan Xie,}
\author[2,3]{Giacomo Cacciapaglia,}
\author[4]{Thomas Flacke}

\affiliation[1]{Center for Theoretical Physics, Department of Physics and Astronomy, Seoul National University, Seoul 08826, Korea}
\affiliation[2]{University of Lyon, Universit{\'e} Claude Bernard Lyon 1, Lyon 69001, France}
\affiliation[3]{Institut de Physique des 2 Infinis de Lyon (IP2I), CNRS/IN2P3 UMR5822, Villeurbanne 69622, France}
\affiliation[4]{Center for Theoretical Physics of the Universe, Institute for Basic Science (IBS), Daejeon 34051, Korea}

\emailAdd{kpxie@snu.ac.kr}
\emailAdd{g.cacciapaglia@ipnl.in2p3.fr}
\emailAdd{flacke@ibs.re.kr}
\abstract{Exotic decays of top partners in  new bosons are the norm in realistic models of a composite Higgs. We focus on the custodial charge-$5/3$ partner, which normally decays exclusively into $tW^+$. The new channels include a colour-sextet, $X_{5/3} \to \bar{b} \pi_6$, as well as singly and doubly charged scalars, $X_{5/3} \to t \phi^+$, $b \phi^{++}$. We use existing same-sign lepton searches to show that the new final states are constrained at the same level as the standard one. At the same time, exotic final states also offer opportunities for improvement: examples include a hard photon in $X_{5/3}\rightarrow t\phi^+\rightarrow tW^+\gamma$ decays, and top-rich channels which arise in several exotic $X_{5/3}$ decays.}

\begin{document}
\hspace*{112mm}{\large \tt CTPU-PTC-19-20} \\

\maketitle
\flushbottom

%%%%%%%%%%%%%%%%%%%%%%%%%%%%%%%%%%%%%%%%%%%%%%%%%%%%%%
%%%%%%%%%%%%%%%%%%%%%%%%%%%%%%%%%%%%%%%%%%%%%%%%%%%%%%

\section{Introduction}

Vector-like quarks (VLQs) commonly emerge as top partners in composite Higgs models where the top acquires a mass via a linear mixing in the partial compositeness mechanism~\cite{Kaplan:1991dc}. The VLQs are thus fermionic bound states transforming as multiplets of the underlying global symmetry that is spontaneously broken in order to generate a composite Higgs as a pseudo-Nambu-Goldstone boson (pNGB). 
Being composites, they couple directly to the pNGB Higgs and they communicate the electroweak symmetry breaking to the elementary top fields.
The electroweak group $SU(2)_L\times U(1)_Y$ is a gauged subgroup of the unbroken global symmetry group, such that the top partner multiplets can be classified in terms of their electroweak quantum numbers. A top partner $X_{5/3}$ with charge $+ 5/3$ always occurs as part of an $SU(2)_L\times SU(2)_R$ bi-doublet in models that preserve custodial symmetry~\cite{Agashe:2006at,Contino:2006qr}.  This situation can also occur in models with extra dimensions (see, e.g., Ref.~\cite{Carena:2006bn}) and in models where VLQ multiplets are added via renormalisable couplings (see, e.g., Refs~\cite{delAguila:2000rc,Buchkremer:2013bha,Cacciapaglia:2015ixa}).

Searches for $X_{5/3}$ constitute an important part of the ATLAS~\cite{Aaboud:2018uek,Aaboud:2018xpj} and CMS~\cite{Sirunyan:2018yun} search programs. The exotically charged state is pair-produced via its QCD interactions and is usually assumed to decay into $tW^+$~\cite{Dennis:2007tv,Contino:2008hi}, which is the only 2-body decay into Standard Model (SM) particles allowed by its quantum numbers. Decays into lighter flavours, namely $cW^+$ or $uW^+$, are also possible~\cite{Atre:2011ae,Cacciapaglia:2011fx,Delaunay:2013pwa,Cacciapaglia:2018lld} and lead to weaker constraints by searches for light quark partners~\cite{Sirunyan:2017lzl,Aad:2015tba}. Furthermore, $X_{5/3}$ can also be singly produced through its interaction with $tW^+$, and this channel has been extensively studied both experimentally~\cite{Sirunyan:2019tib,Aaboud:2018xpj} and phenomenologically~\cite{DeSimone:2012fs,Mrazek:2009yu,Azatov:2013hya,Backovic:2014uma,Matsedonskyi:2014mna}. In this work we focus on QCD pair production, which has the benefit of model-independence as the cross-section only depends on the VLQ mass. 
However, other decays of $X_{5/3}$ are possible, especially if the model contains non-SM states that are lighter than the $X_{5/3}$. A classification of the possible final states has been first attempted in Ref.~\cite{Brooijmans:2016vro}. In realistic models of a composite Higgs based on gauge-fermion underlying interactions~\cite{Ferretti:2013kya}, such lighter states always occur in the form of additional pNGBs beyond the Higgs multiplet~\cite{Ferretti:2016upr,Cacciapaglia:2015eqa,Agugliaro:2018vsu}. In Ref. \cite{Bizot:2018tds} a first survey of exotic top partner decays that commonly occur in underlying models has been presented. These include the following new decay channels for $X_{5/3}$:

\begin{enumerate}
	\item {\it Decay to a coloured pNGB $\pi_6$ with charge 4/3: $X_{5/3}\rightarrow \bar{b}\pi_6\rightarrow \bar{b} tt$.}\\
	The colour-sextet $\pi_6$ emerges in models with an $SU(6)/SO(6)$ breaking pattern in the QCD sector.~\footnote{In the models of Ref.~\cite{Ferretti:2013kya}, QCD charges are sequestered to a second species of underlying fermions that transform under a different representation than that of the Higgs constituents. There exist 3 classes of symmetry patterns: $SU(6)/SO(6)$ containing a neutral colour octet and a charge-$4/3$ (or $-2/3$) sextet pNGB, $SU(6)/Sp(6)$ containing a neutral colour octet and a charge-$2/3$ triplet, and $SU(3)^2/SU(3)$ containing only a neutral octet~\cite{Ferretti:2016upr}.} It decays exclusively into two same-sign tops~\cite{Cacciapaglia:2015eqa}. 
	\item {\it Decay to electroweakly charged pNGBs: $X_{5/3}\rightarrow t\phi^+$ and  $X_{5/3}\rightarrow b\phi^{++}$.}\\
	Singly and doubly charged scalars are present, for instance, in models with breaking patterns $SU(5)/SO(5)$~\cite{Dugan:1984hq,Agugliaro:2018vsu} and $SU(6)/SO(6)$~\cite{Cacciapaglia:2019ixa} in the electroweak sector. They emerge as a bi-triplet of $SU(2)_L\times SU(2)_R$, i.e. a charged and a neutral $SU(2)_L$ triplet.~\footnote{Singly-charged pNGBs also arise in the symmetry patterns $SU(4)^2/SU(4)$~\cite{Ma:2015gra} and $SU(6)/Sp(6)$~\cite{Low:2002ws,Cai:2018tet}, both of which enjoying underlying gauge-fermion descriptions.} The 9 degrees of freedom can be classified in terms of the diagonal $SU(2)$ preserved by the Higgs~\cite{Dugan:1984hq}, and they form a custodial quintuplet $\bf{5}= (\phi^{++}_5, \phi^{+}_5,\phi^0_5,\phi^{-}_5,\phi^{--}_5)$ that contains both doubly- and singly-charged scalars, a custodial triplet $\bf{3}= (\phi^{+}_3, \phi^0_3,\phi^{-}_3)$ that contains a second singly-charged scalar, and a singlet. Following the analysis in Ref.~\cite{Agugliaro:2018vsu}, we  identify the following decay patterns for the charged pNGBs:
	\begin{equation}
		(a_1)\;\; \phi^{++}\rightarrow W^+W^+\,, \quad (a_2)\;\; \phi^{++}\rightarrow W^+\phi^+\,, \quad (a_3)\;\; \phi^+\rightarrow W^+\gamma + W^+Z\,, \nonumber
	\end{equation}
	coming from gauge interactions and the WZW topological term;
	\begin{equation}
		(b_1)\;\; \phi^+\rightarrow t\bar{b}\,, \qquad  (b_2)\;\; \phi^+\rightarrow \tau^+\nu\,, \nonumber
	\end{equation}
	generated by partial compositeness for the 3$^{\rm rd}$ generation; ~\footnote{We consider couplings to 3$^{\rm rd}$ generation because the couplings are typically proportional to the fermion masses. For the same reason, we would expect the coupling to top-bottom to dominate over the coupling to leptons.}
	\begin{equation}
		(c_1)\;\; \phi^{++}\rightarrow \tau^+\tau^+\,, \qquad (c_2)\;\; \phi^+\rightarrow \tau^+\bar{\nu}\,, \nonumber
	\end{equation}
	generated by lepton violating interactions ($\Delta L=2$), that may be related to neutrino mass generation.~\footnote{Conservatively, we allow such couplings to be sizeable.} Typically, as the couplings are of very different origins, we would expect one set of decays to dominate over the others, barring tuned values of the parameters.
	
\end{enumerate}

\begin{table}[t]
	\footnotesize\renewcommand\arraystretch{1.5}\centering
	\begin{tabular}{|c|c|c|c|}\hline
		\multicolumn{3}{|c|}{Cascade decays} &after $t$ and $\tau$ decay \\ \hline
		\multirow{8}{*}{$X_{5/3}$} & $tW^+$ & $-$ & $(bW^+)W^+$ \\ \cline{2-4}
		& $\bar b\pi_6$ & $\bar b tt$ &  $\bar{b}(bW^+)(bW^+)$\\  \cline{2-4}
		& \multirow{3}{*}{$t\phi^+$} & $tW^+\gamma$, $tW^+Z$ & $(bW^+)W^+\gamma$, $(bW^+)W^+Z$\\  \cline{3-3}\cline{4-4}
		& & $tt\bar b$ &  $(bW^+)(bW^+)\bar b$ \\   \cline{3-3}\cline{4-4} 
		& & $t\tau^+\nu$ & $(bW^+)(W^{+*}\bar{\nu})\nu$ \\  \cline{2-4}
		& \multirow{3}{*}{$b\phi^{++}$} & $bW^+W^+$ & $bW^+W^+$ \\ \cline{3-3}\cline{4-4}
		& &  $bW^{+(*)}\phi^{+}$ &  $bW^{+(*)}W^{+(*)}+X$  \\  \cline{3-3}\cline{4-4}
		& & $b\tau^+\tau^+$ &  $b(W^{+(*)}\bar{\nu})(W^{+(*)}\bar{\nu})$ \\ \hline
	\end{tabular}
	\caption{Possible decay channels of $X_{5/3}$. In the right-most column, we indicate the final state after $t$ and $\tau$ decays in order to underline their similarity. $W^*$ denotes off-shell $W$ bosons while $W^{(*)}$ denotes $W$ bosons which are on- or off-shell, depending on the mass spectrum. The decay products of $\phi^+$ always contain one $W^{+(*)}$, and we  label the final state of the decay chain $X_{5/3}\to b\phi^{++}\to bW^{+(*)}\phi^{+}$ as $bW^{+(*)}W^{+(*)}+X$, where $X$ denotes  $\gamma$, $Z$, $b\bar{b}$, or $\bar{\nu}\nu$.}
	\label{tab:decay_channels_schem}
\end{table}

Coloured or electroweakly charged pNGBs add a plethora of new decay channels for $X_{5/3}$, which differ in the number of tops, taus and on-shell $W$ bosons. However, unless specific tags for $t$, $\tau$ or $W$ are explicitly added in the search, the resulting final states share many similarities, as summarised in Table~\ref{tab:decay_channels_schem}. The similarity becomes even more clear when writing the decay products after $t\rightarrow bW^+$ and $\tau^+\rightarrow W^{+*}\bar{\nu}$ decays, where $W^*$ denotes an off-shell $W$ boson.  All final states contain at least one $b$-jet as well as two same-sign $W$ bosons, either on- or off-shell.
Existing $X_{5/3}$ searches focus on two complementary strategies: either they tag same-sign leptons (SSL)~\cite{Sirunyan:2018yun,Aaboud:2018xpj} originating from a fully leptonic decay chain $X_{5/3}\rightarrow t_{\rm{lep}} W_{\rm{lep}}^+$, or they require a single lepton plus jets (SLj)~\cite{Aaboud:2018uek,Sirunyan:2018yun} thus targeting the semi-leptonic decay chain $X_{5/3}\rightarrow t_{\rm{lep}} W_{\rm{had}}^+$ or $t_{\rm{had}} W_{\rm{lep}}^+$. 
Assuming $\Br(X_{5/3}\rightarrow tW^+)=1$, the current bound on the mass $M_{X_{5/3}}$ from the SSL channel is slightly weaker than that from the SLj channel. More concretely, according to the latest results at 35.9 fb$^{-1}$~\cite{Sirunyan:2018yun}, the observed lower limit for the SSL and SLj channels are 1.16 TeV and 1.30 TeV respectively, for a right-handed $\bar X_{5/3}-t-W^+$ coupling. However, the expected (i.e.  Monte Carlo) bounds are rather similar for those two channels: 1.19 TeV for the SSL and 1.23 TeV for SLj, implying that the two channels have almost the same sensitivity to the $X_{5/3}\bar X_{5/3}\to tW^+\bar tW^-$ final state.

Both search strategies apply to the exotic $X_{5/3}$ decays, however the kinematics and the particle multiplicities are modified. 
The SSL channel has a few advantages compared to the SLj one:
it has very low SM background and searches impose few specialised cuts beyond demanding high-$p_T$ same-sign leptons (and two or more jets). We thus expect the SSL searches to be more sensitive to most of the exotic decay modes. 
In this work, we study how the bounds on the $X_{5/3}$ mass are modified in presence of the new decay channels, which are theoretically motivated by realistic composite Higgs models. Furthermore, we will identify new requirements that can be added to current searches to significantly improve the reach in some specific cases. The latter is a great opportunity to improve the performance of the LHC searches in view of the high-luminosity phase that will start in a few years.

The article is structured as follows: In Section \ref{sec.models} we provide simplified models featuring the decays $X_{5/3}\rightarrow \bar{b}\pi_6\rightarrow \bar{b} tt$, $X_{5/3}\rightarrow t\phi^+$ and $X_{5/3}\rightarrow b\phi^{++}$, and the subsequent decays of the scalars, motivated by the underlying models of Ref.~\cite{Bizot:2018tds}.
In Section \ref{sec.SSL}, we recast the $X_{5/3}$ search in the SSL channel  \cite{Sirunyan:2018yun}, determine the bounds that apply if exotic $X_{5/3}$ decays are present, and provide projections for the exclusion reach of the high-luminosity LHC run (HL-LHC). 
The recast search can also be applied to multi top final states, as arising from the decays of charge-$2/3$ top partners via a singlet (pseudo-)scalar, $T \to t a \to t t \bar{t}$ as well as $T \to t a \to t W^+ W^-$, and we show the results in Appendix~\ref{app:Tta}.
In Section \ref{sec.SSLgam}, we investigate the opportunities for probing the exotic decays of $X_{5/3}$, such as adding a new hard photon requirement that can improve the reach on the decay $X_{5/3}\rightarrow t\phi^{+}\rightarrow tW^+\gamma$ in the SSL plus photon channel, due to very low SM backgrounds. In addition, the collinearity of the SSLs and the number of jets/$b$-jets are also useful to distinguish exotic decays from the standard one. 
We conclude in Section \ref{sec.conclusions}.

%%%%%%%%%%%%%%%%%%%%%%%%%%%%%%%%%%%%%%%%%%%%%%%%%%%%%%
%%%%%%%%%%%%%%%%%%%%%%%%%%%%%%%%%%%%%%%%%%%%%%%%%%%%%%

\section{Simplified Models} \label{sec.models}

\subsection{$X_{5/3}\rightarrow \bar{b}\pi_6\rightarrow \bar{b} tt$} \label{sec.pi6mod}

As a first example, we consider exotic decays of $X_{5/3}$ in the presence of a colour-sextet pseudo-scalar, which for example occurs in underlying models of top partial compositeness with $SU(6)/SO(6)$ breaking in the colour sector~\cite{Cacciapaglia:2015eqa}. 
In this case, the sextet $\pi_6$ emerges as a pNGB and is a singlet of $SU(2)_L$ with charge $4/3$.

The effective Lagrangian for the $X_{5/3}$ couplings, including the sextet, reads  \cite{Cacciapaglia:2015eqa,Bizot:2018tds}
\begin{multline}
	\mathcal{L}_{X_{5/3}}^{\pi_6} =\bar{X}_{5/3} \left(i \slashed{D} - M_{X_{5/3}}\right)X_{5/3} 
	\label{eq:XpiLag}\\ 
	+ \left(\kappa^X_{W,R} \frac{g}{\sqrt{2}} \, \bar{X}_{5/3} \slashed{W}^+P_R t  + i \kappa^{X}_{\pi_6,L}  \, \bar{X}_{5/3} \pi_6 (P_L b)^c  + L\leftrightarrow R + \mbox{h.c.} \right)~,
\end{multline}
where the superscript ``$c$'', as in $b^c$, denotes the charge conjugate of the field. For the sake of generality, we allow both chirality structures for the couplings of $X_{5/3}$ with the sextet, however in composite Higgs models it is the coupling with a left-handed bottom that dominates. This coupling is in fact related to the degree of compositeness of the left-handed doublet (which contains also the left-handed top), which is required to be of order 1 to generate a large enough top mass. On the other hand, the coupling to the right-handed bottom is suppressed by the small compositeness of the bottom, thus being smaller by a factor of $m_b/m_t$ than the other chirality coupling. In addition to this, in models within the classification of Ref.~\cite{Ferretti:2013kya} the $X_{5/3}$ belongs to a $SU(2)_L$ doublet while the $\pi_6$ is a singlet, thus the right-handed coupling is suppressed by an additional factor of $v/f \ll 1$ with respect to the left-handed coupling ($v$ is the electroweak scale while $f$ is the decay constant of the pNGBs). The latter suppression can, however, be changed in other models where the composite states belong to a different multiplet of $SU(2)_L$: for instance, if the $\pi_6$ belonged to an $SU(2)_L$ doublet, the $v/f$ factor would appear in the left-handed coupling, thus it could balance the $m_b/m_t$ suppression. The same case would occur if $X_{5/3}$ were a singlet. For simplicity, however, in the following we will only consider couplings to the left-handed bottom. Regarding the coupling to the $tW^+$, it is always generated by a mixing proportional to $v$, thus the dominant chirality will only depend on the $SU(2)_L$ representation the $X_{5/3}$ belongs to: right-handed top for a doublet and left-handed top for a singlet.

The $X_{5/3}$ branching ratios depend on the size of the effective couplings $ \kappa^{X}_{\pi_6,L/R}$ and $\kappa^X_{W,L/R}$ as well as the mass ratios $M_{X_{5/3}}:M_{\pi_6}:m_W$. 
In underlying models, various branching ratios can be realized, including dominance of $X_{5/3}\rightarrow tW^+$, dominance of $X_{5/3}\rightarrow \bar{b}\pi_6$, and comparable branching fractions~\cite{Cacciapaglia:2015eqa}. For the phenomenological study performed in this article, we thus treat the branching fraction as a free parameter. 

The effective Lagrangian for the $\pi_6$  couplings to SM particles is \cite{Cacciapaglia:2015eqa,Bizot:2018tds}
\bea\label{eq:XpiLag1}
\mathcal{L}_{\pi_6} =
\left| D_\mu \pi_6 \right|^2 - M_{\pi_6}^2 \left|\pi_6 \right|^2 + \left( i \kappa^{\pi_6}_{tt,R} \,   \bar{t} \pi_6 (P_R t)^c + L\leftrightarrow R + \mbox{h.c.}\right)~,
\eea
where $t^c$ denotes the charge conjugate of the  the top quark fields.  In the underlying models with an $SU(2)$ singlet $\pi_6$, the coupling $\kappa^{\pi_6}_{tt,L}$ is suppressed by $m_t^2/f_{\pi_6}^2$ with respect to $\kappa^{\pi_6}_{tt,R}$, and the sextet decays as $\pi_6 \to tt$, with large dominance to right-handed tops \cite{Cacciapaglia:2015eqa}.

\subsection{$X_{5/3}\rightarrow t\phi^+$ and $X_{5/3}\rightarrow b\phi^{++}$} \label{sec.chmod}

As a second example, we consider exotic decays of $X_{5/3}$ in the presence of electroweakly charged pNGBs, which for example are present in underlying models with $SU(5)/SO(5)$ breaking in the electroweak sector \cite{Agugliaro:2018vsu}.

The effective Lagrangian for the VLQ $X_{5/3}$ and the charged scalar couplings reads \cite{Bizot:2018tds}
\bea\label{eq:XphiLag}
\mathcal{L}_{X_{5/3}}&=&
\bar{X}_{5/3} \left(i \slashed{D} - M_{X_{5/3}}\right)X_{5/3} 
+\left(\kappa^X_{W,R} \frac{g}{\sqrt{2}} \, \bar{X}_{5/3} \slashed{W}^+ P_R t+L \leftrightarrow R +\hc\right)\nn\\
&&+\left( i \kappa^{X}_{\phi^+,L}  \, \bar{X}_{5/3} \phi^+ P_L t 
+ i \kappa^{X}_{\phi^{++},L}  \, \bar{X}_{5/3} \phi^{++} P_L b + L \leftrightarrow R +\hc\right).
\eea
Again, the $SU(2)_L$ quantum numbers of $X_{5/3}$, $\phi^+$, and $\phi^{++}$ imply dominance of one chirality in the couplings : for $X_{5/3}$ belonging to a doublet and $\phi^{+,++}$ coming from triplets, the dominant couplings are $\kappa^X_{W,R}$, $\kappa^{X}_{\phi^+,L}$, and $\kappa^{X}_{\phi^{++},L}$, with the others suppressed by an additional $v/f$. Moreover, the coupling involving a right-handed bottom $\kappa^{X}_{\phi^{++},R}$ is suppressed by the bottom Yukawa.

Concerning the interactions among the light scalars and the SM particles, we have
\bea\label{eq:Lphi}
\mathcal{L}_{\phi^+}&=&\left|D_\mu \phi^+ \right|^2 - M_{\phi^+}^2 \left|\phi^+ \right|^2
+\left(\frac{g^2c_W \kappa^\phi_{WB}}{8\pi^2 f_\phi} \phi^+W^-_{\mu\nu}\tilde{A}^{\mu\nu}
-\frac{eg \kappa^\phi_{WB}}{8\pi^2 f_\phi} \phi^+W^-_{\mu\nu}\tilde{Z}^{\mu\nu}+\hc\right)\nonumber \\
&&+\left( i \kappa^\phi_{tb,L} \frac{m_t}{f_\phi} \,  \bar{t} \phi^+ P_L b + L\leftrightarrow R +\hc \right)
+\left( i \kappa^\phi_{\tau\nu,R} \frac{m_\tau}{f_\phi} \,  \bar\nu \phi^+ P_R \tau + \hc\right)\,,
\eea
for the singly-charged scalar. In underlying models, the interactions in the first line result from anomalous couplings of the bound state $\phi^+$ to the standard model gauge bosons, and the anomaly coefficients are determined by the SM quantum numbers of the electroweakly charged constituent fermions of $\phi^+$~\cite{Dugan:1984hq}. Note that the $\phi^+\to W^+\phi^0$ chain decay is typically subdominant compared to the WZW induced decays~\cite{Agugliaro:2018vsu}. The structure of the WZW couplings derives from the fact that $\phi^+$ transforms as a $SU(2)_L$ triplet, thus there exists an unsuppressed $\phi^+- W^--B$ coupling ($B$ being the hypercharge gauge boson), which results in 
\bea
\Br(\phi^+\to W^+\gamma):\Br(\phi^+\to W^+Z)=\cos^2 \theta_W: \sin^2 \theta_W\approx80\%:20\%,
\eea
when $M_{\phi^+}\gg m_W$. The couplings of $\phi^+$ to SM fermions in the second line of Eq.~(\ref{eq:Lphi}) typically arise out of the mechanism generating SM fermion masses, thus explaining the proportionality to the quark and lepton masses. Note that contrary to the WZW couplings, the couplings to fermions may vanish for some choices of the VLQ representations entering the partial compositeness sector~\cite{Agugliaro:2018vsu}. Furthermore, such couplings, if present, typically dominate over the WZW couplings, the former being generated at tree level while the latter are loop suppressed. We can thus identify three distinct scenarios:
\begin{enumerate}
	
	\item Couplings to fermions vanish, thus the decays into a pair of gauge bosons dominate, $\phi^+ \to W^+ \gamma+W^+Z$;
	
	\item The coupling to $3^{\rm rd}$ generation quarks is present, thus the dominant decay is $\phi^+ \to t\bar b$ (left-handed bottom);
	
	\item The coupling to quarks vanish while only the coupling to leptons is present, thus the dominant decay is $\phi^+ \to \tau^+\nu$.
	
\end{enumerate}
For the phenomenology, therefore, we will only consider these 3 cases. Note that in principle we could add a coupling to the left-handed tau, however this would imply the presence of a right-handed neutrino and the coupling would be very small because it is proportional to the neutrino masses.

\bigskip

For the doubly charged scalar, we have
\bea\label{eq:XphiLag2}
\mathcal{L}_{\phi^{++}}&=&\left|D_\mu \phi ^{++}\right|^2 - M_{\phi^{++}}^2 \left|\phi^{++} \right|^2
+\left(\frac{g^2 \kappa^\phi_{WW}}{8 \pi^2 f_\phi} \phi^{++} W^-_{\mu\nu} \tilde{W}^{\mu\nu,-} +\hc\right)\,, 
\eea
where the only allowed coupling mediating decays comes from the WZW interactions. As $\phi^{++}$ is a member of a charged triplet, this coupling must necessarily be suppressed by powers of the Higgs scale, namely $v^2/f^2$, thus it is very small. This leaves open the possibility of sizeable chain decays 
$\phi^{++}\to W^+\phi^{+}$  arising from the gauge interaction term 
\be
\mathcal{L}_{\phi^{++}}\supset g\kappa_{W\phi}^\phi(\phi^-\partial^\mu\phi^{++}-\partial^\mu\phi^-\phi^{++})W_\mu^-\,.
\ee
Due to the suppression in the WZW coupling, the chain decay may dominate even if the two charged scalars are very close in mass (as they belong to the same multiplet) and the $W$ is off-shell~\cite{Agugliaro:2018vsu}. Even though in the $SU(5)/SO(5)$ model there are two singly-charged states, it turns out that, in realistic configurations, their masses are always very close.

There is another scenario that leads to interesting phenomenology if the model contains sources of lepton number violation, which generate the following $\Delta L = 2$ couplings:
\bea
\mathcal{L} \supset  \left( i \tilde{\kappa}^\phi_{\nu\tau,L} \,  \bar{\nu^c} \phi^{+} P_L\tau + \hc\right) +  \left( i \tilde{\kappa}^\phi_{\tau\tau,L} \,  \bar{\tau^c} \phi^{++} P_L\tau + L\leftrightarrow R + \hc\right)\,,
\eea
where the coupling to right-handed taus are suppressed by $v^2/f^2$. These terms may be related to Majorana neutrino mass generation, thus being very small, however here we will be pragmatic and allow them to dominate. They add an interesting leptonic decay channel for the doubly-charged scalar, while the decay of the singly-charged scalar would be indistinguishable from the one in Eq.~\eqref{eq:Lphi}.

\begin{table}[h]
	\footnotesize\renewcommand\arraystretch{1.5}\centering
	\begin{tabular}{|c|c|c|c|c|c|c|c|c|c|}\hline
		\multicolumn{3}{|c|}{Cascade decays} & \multicolumn{2}{|c|}{Relevant couplings} & Relevant masses \\ \hline
		\multirow{8}{*}{$X_{5/3}$} & $tW^+$ & $-$ & \multicolumn{2}{|c|}{$\kappa_{W,L/R}^X$} & $M_{X_{5/3}}$\\ \cline{2-6}
		& $\bar b\pi_6$ & $\bar b tt$ & $\kappa_{\pi_6,L/R}^X$ & $\kappa^{\pi_6}_{tt,L/R}$ & $M_{X_{5/3}}$, $M_{\pi_6}$ \\ \cline{2-6}
		& \multirow{3}{*}{$t\phi^+$} & $tW^+Z$, $tW^+\gamma$ & \multirow{3}{*}{$\kappa_{\phi^+,L/R}^X$} & $\kappa_{WB}^{\phi}$ & \multirow{3}{*}{$M_{X_{5/3}}$, $M_{\phi^+}$}\\ \cline{3-3}\cline{5-5}
		& & $tt\bar b$ & & $\kappa_{tb,L/R}^{\phi}$ & \\  \cline{3-3}\cline{5-5}
		& & $t\tau^+\nu$ & & $\kappa_{\tau\nu,L/R}^{\phi}$ & \\ \cline{2-6}
		& \multirow{3}{*}{$b\phi^{++}$} & $bW^+W^+$ & \multirow{3}{*}{$\kappa_{\phi^{++},L/R}^X$} & $\kappa_{WW}^{\phi}$ & $M_{X_{5/3}}$, $M_{\phi^{++}}$\\ \cline{3-3}\cline{5-6}
		& &  $bW^{+(*)}\phi^{+}$ & & $\kappa_{W\phi}^{\phi}$ & $M_{X_{5/3}}$, $M_{\phi^{++}},M_{\phi^{+}}$\\  \cline{3-3}\cline{5-6}
		& & $b\tau^+\tau^+$ & & $\kappa_{\tau\tau,L/R}^{\phi}$ & $M_{X_{5/3}}$, $M_{\phi^{++}}$\\ \hline
	\end{tabular}
	\caption{Decay channels of $X_{5/3}$. For each channel, we indicate the relevant couplings and BSM masses.} 
	\label{tab:decay_channels}
\end{table}

The different decay channels of $\phi^{++}$ and $\phi^{+}$ imply a large number of possible final states from $X_{5/3}$ decays which are summarised in Table~\ref{tab:decay_channels}, together with the relevant couplings and particle masses. For completeness, we also include the decays from the scenario described in Sec. \ref{sec.pi6mod}.

%%%%%%%%%%%%%%%%%%%%%%%%%%%%%%%%%%%%%%%%%%%%%%%%%%%%%%
%%%%%%%%%%%%%%%%%%%%%%%%%%%%%%%%%%%%%%%%%%%%%%%%%%%%%%

\section{Bounds: same-sign leptons}\label{sec.SSL}

In this section we extract bounds on various decay modes of QCD pair produced $X_{5/3}$. We focus on the SSL final state, which is common to all decay modes, and recast the CMS search of Ref.~\cite{Sirunyan:2018yun}, which currently provides the strongest bound on the standard $X_{5/3}$ decay among the existing SSL searches~\cite{Aaboud:2018xpj,Sirunyan:2018yun}.
The recast and its validation are described in Sec.~\ref{sec.recast}, while in the following Sections~\ref{sec:pi6} - \ref{sec:phipp} we determine the signal efficiencies for the final states deriving from the new decay modes of $X_{5/3}$. We determine bounds on the mass of $X_{5/3}$ assuming that both of the pair-produced $X_{5/3}$ decay in the same way.

As we will see, all exclusive decay modes have very similar efficiencies, thus we expect that events with mixed decay modes also have similar efficiencies. This is mainly due to the fact that the SSL pair is present in all final states, and the only difference is in the kinematics of the SM particles in the final state.
The SSL final state is also relevant for other decays that are rich in tops, like for instance the 6-top final state deriving from the decays of a charge-$2/3$ top partner via a neutral (pseudo-)scalar, $T \to t a \to t t \bar{t}$~\cite{Bizot:2018tds}. This final state has been already studied in Refs~\cite{Deandrea:2014raa,Han:2018hcu}, and as a by-product of our study we applied our recast search to this channel and find the strongest bound on the $T$ mass of $1.3$~TeV (details are shown in Appendix~\ref{app:Tta}).

\subsection{Recast of the CMS SSL search}\label{sec.recast}

We write the Lagrangians described in Section~\ref{sec.models}  in model files using the {\tt FeynRules} package~\cite{Alloul:2013bka} to implement them in {\tt MadGraph5\_aMC@NLO}~\cite{Alwall:2014hca} which is used to generate the parton-level signal events. Generated events are matched to $+1{\rm~jet}$ final state and then interfaced to {\tt Pythia 8}~\cite{Sjostrand:2014zea} and {\tt Delphes}~\cite{deFavereau:2013fsa} for parton shower and fast detector simulation, respectively. The events are generated at LO accuracy in QCD, however, with the help of the {\tt Top++2.0} package~\cite{Czakon:2011xx,Czakon:2013goa,Czakon:2012pz,Czakon:2012zr,Baernreuther:2012ws,Cacciari:2011hy}, we normalise the $pp\to X_{5/3}\bar X_{5/3}$ cross-section with the QCD next-to-next-to-leading order (NNLO) one with next-to-next-to-leading logarithm (NNLL) soft-gluon resummation. 

We then implement a cut-flow that mimics the search strategy of Ref.~\cite{Sirunyan:2018yun}, namely events are required to have:
\begin{enumerate}
	\item At least two SSLs with the leading one $p_T^{\ell}>40$~GeV and the sub-leading one $p_T^\ell>30$~GeV. The invariant mass of the SSL pair is required to be larger than $20$~GeV to veto quarkonium resonances. For the purpose of suppressing the $Z\to\ell^+\ell^-$ background, events containing a pair of opposite-sign and same-flavour leptons whose invariant mass is within $15$~GeV of the $Z$ boson mass are rejected. For the events with two or more electrons, the veto applies to the $e^\pm e^\pm$ pair as well, to suppress backgrounds from charge-misidentified electrons. 
	\item At least two AK4 jets with $p_T^j>30$~GeV and $|\eta^j|<2.4$. The AK4 jets are reconstructed by the anti-$k_t$ algorithm with a cone size $\Delta R=0.4$.
	\item We require a minimum number of constituents $N_{\rm const}\geqslant5$, where $N_{\rm const}=N_j+N_\ell-2$ with $N_j$ and $N_\ell$ being the number of AK4 jets and charged leptons, respectively.
	\item $H_T>$ 1.2 TeV, where $H_T$ is the scalar sum of the transverse momenta of all reconstructed objects.
\end{enumerate}
\noindent For the background, we take the expected (i.e. Monte Carlo) numbers given in the CMS study  Ref.~\cite{Sirunyan:2018yun}, which are $10.9\pm 1.9$ events with a  same-sign di-electron, $11.2\pm 2.0$ with di-muons and $23.2\pm 3.7$ with an electron-muon pair.

\begin{figure}[htbp]
	\begin{center}
		\includegraphics[width=0.45\textwidth]{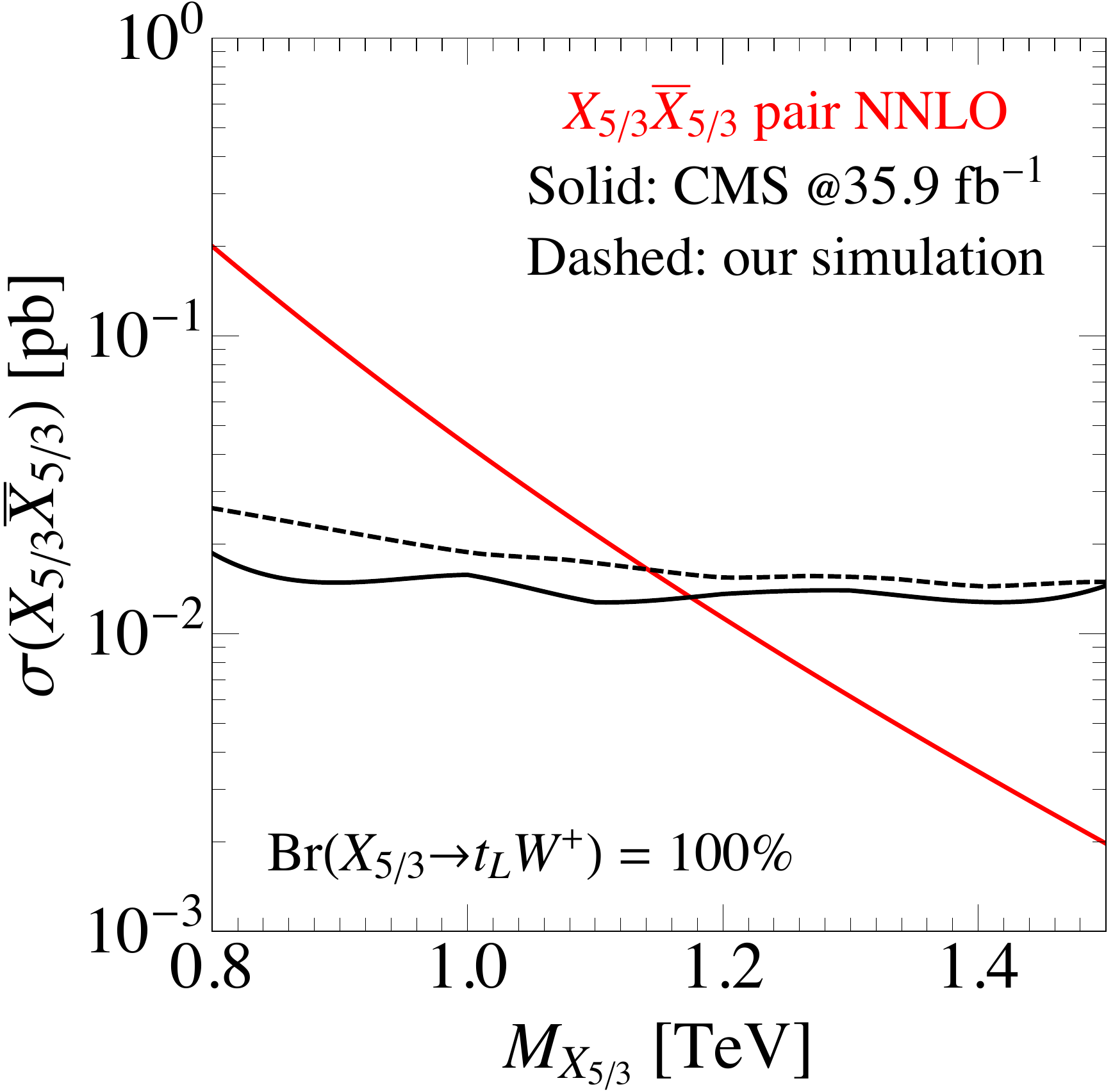}\qquad
		\includegraphics[width=0.45\textwidth]{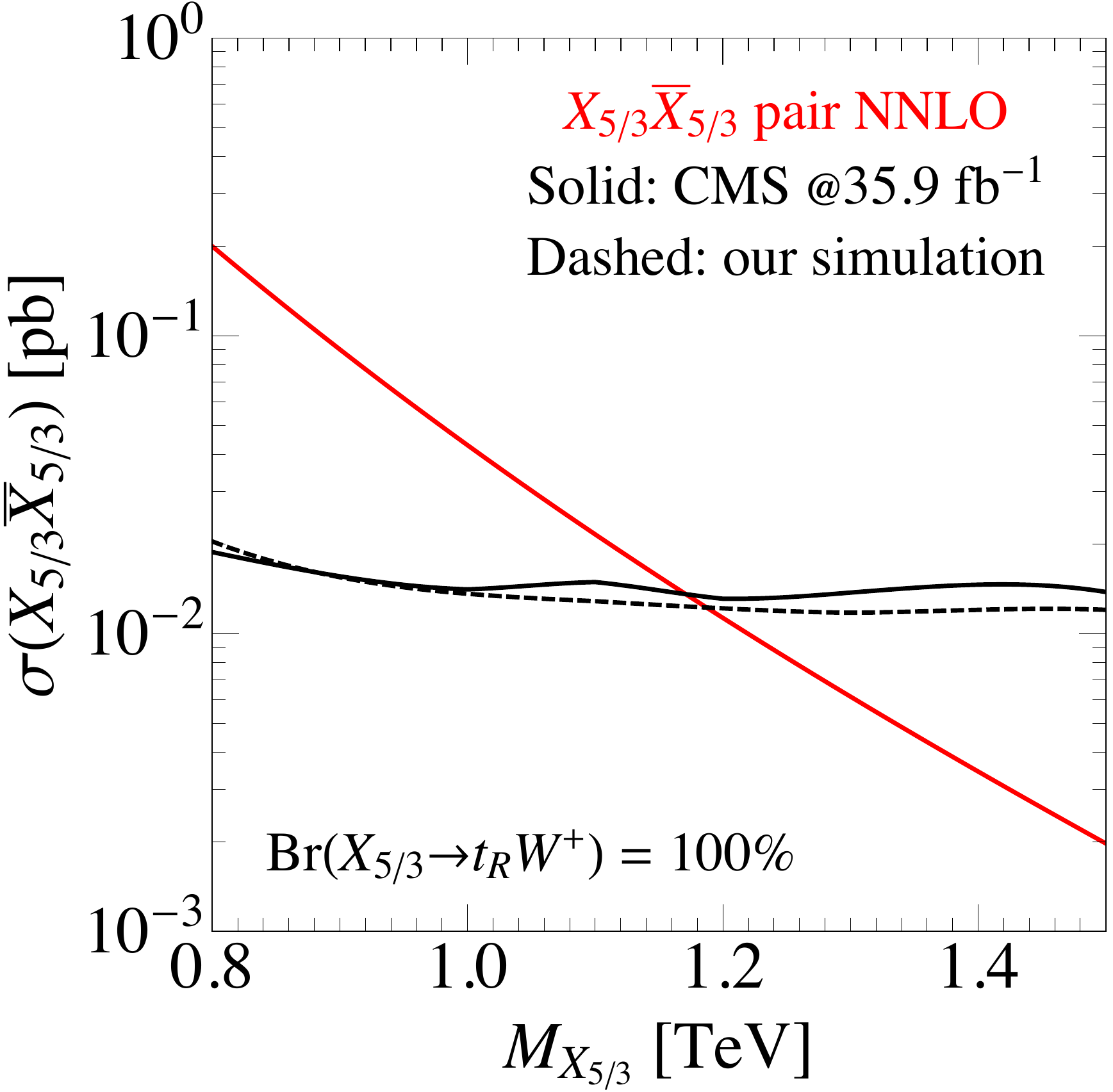}
		\caption{Expected bound on the QCD $X_{5/3}$ pair production cross-section from our recast (dashed) compared to the CMS results~\cite{Sirunyan:2018yun} (solid) for left-handed (left) and right-handed (right) couplings. In red we show the $X_{5/3}$ pair production cross section at NNLO-NNLL.}
		\label{fig.recast}
	\end{center}
\end{figure}

To validate our recast, we generate events with pair production of $X_{5/3}$ with subsequent decay $X_{5/3}\rightarrow tW$ both via a right-handed  and a left-handed coupling, and compare the bound on the cross section we obtain with the official CMS one, as shown in Fig.~\ref{fig.recast}. The results feature  an excellent agreement with the experimental results. In the limit $\Br(X_{5/3}\rightarrow t_RW)=100\%$, the measured (expected) SSL limit on $M_{X_{5/3}}$ is  $1.16$~TeV ($1.19$~TeV) according to CMS~\cite{Sirunyan:2018yun}. To obtain a naive estimate for the HL-LHC reach with a luminosity of 3 ab$^{-1}$ at $13$ TeV, we rescale signal and background event numbers according to the increased luminosity and assume an improvement of the sensitivity by a factor $\mathcal{S}/\sqrt{\mathcal{B}}$. This procedure yields a projected exclusion reach of $M_{X_{5/3}} = 1.56$~TeV. 
We will use the same procedure to extract projected sensitivities for the exotic decays in the next sections.

\subsection{Bounds for $X_{5/3}\rightarrow \bar{b}\pi_6\rightarrow \bar{b} tt$}\label{sec:pi6}

Let us start with the model described in Section~\ref{sec.pi6mod} where the $X_{5/3}$ VLQ is accompanied by a colour-sextet scalar $\pi_6$ with charge $4/3$, which decays to a pair of same-sign  tops, $\pi_6 \to tt$. If the sextet is lighter than the $X_{5/3}$, the two kinematically allowed decays of the VLQ within this model are the standard decay, $X_{5/3}\rightarrow tW^+$, and the exotic decay, $X_{5/3}\rightarrow \bar{b}\pi_6\rightarrow \bar{b} tt$. 
Both decay chains yield a SSL final state with the same probability,
\bea
X_{5/3} & \to & t W^+ \to b W^+_{\rm lep} W_{\rm lep}^+\,, \nonumber \\
X_{5/3} & \to & \bar{b} t t \to \bar{b} b b W^+_{\rm lep} W^+_{\rm lep}\,,  \label{eq:Xbtt}
\eea
such that differences in acceptance result solely from different cut-efficiencies. In the following we will consider only the chirality structure of the couplings that derives from realistic composite models, i.e. $X_{5/3}$ decays dominantly into a right-handed top (plus $W^+$) and a right-handed anti-bottom (via $\pi_6$), while the sextet decays into right-handed tops.

\begin{figure}[htbp]
	\begin{center}
		\includegraphics[width=0.45\textwidth]{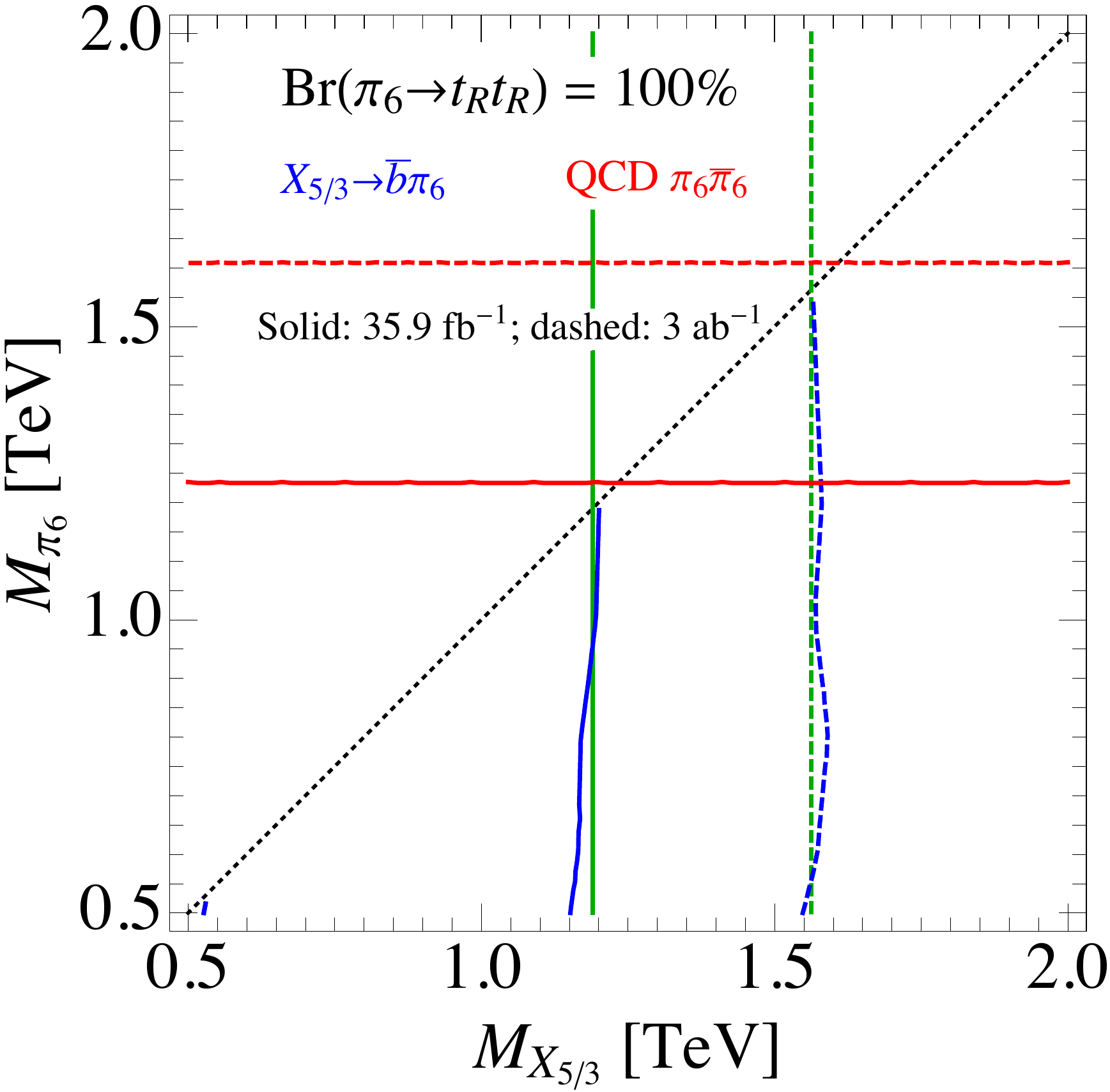}
		\caption{Bounds on the $M_{X_{5/3}}$--$M_{\pi_6}$ plane from the CMS SSL search of Ref.~\cite{Sirunyan:2018yun} (solid lines) and projections for the HL-LHC reach (dashed lines) for the exotic decay $X_{5/3}\to\bar b\pi_6$ (blue) and the QCD pair produced sextets (red). The bounds and projections for the standard ${\rm Br}(X_{5/3}\to tW^+)=1$ decay are shown for reference in green. The horizontal red lines show the bound from the direct QCD pair production of the sextets derived from the same SSL search~\cite{Sirunyan:2018yun}. The bounds become insensitive to $M_{\pi_6}$ for light masses above threshold.}
		\label{fig.pi6}
	\end{center}
\end{figure}

We determine the analogous limits for the case $\Br(X_{5/3}\rightarrow \bar{b}\pi_6)=100\%$, i.e. in the case in which both of the pair-produced $X_{5/3}$ decay through the colour-sextet. The signal efficiency is a function of $M_{X_{5/3}}$ and $M_{\pi_6}$. As compared to $X_{5/3}\rightarrow tW^+$, $X_{5/3}\rightarrow \bar{b} tt$ yields more $b$ jets in the final state. The SSLs result from leptonic decays of the $\pi_6\rightarrow tt$ and tend to be softer and have a smaller angular separation for lighter $\pi_6$. However, the SSL analysis \cite{Sirunyan:2018yun} is a cut-and-count analysis with low background. The cuts are conservative, and the modified kinematics only weakly affect the signal efficiency. Fig.~\ref{fig.pi6} shows the resulting limit (blue solid line) and the projected HL-LHC exclusion reach (blue dashed line). The limits for $X_{5/3}\rightarrow \bar{b}\pi_6\rightarrow \bar{b} tt$ weakly depend on $M_{\pi_6}$ and are comparable to the limits for $X_{5/3}\rightarrow tW^+$ (green solid and dashed lines).

The SSL search we recast can also be used to constrain the direct pair production of the sextet via QCD, thus providing a bound on $M_{\pi_6}$ directly~\cite{Cacciapaglia:2015eqa}. The $\pi_6$ decays to $tt$ (and $\pi_6^\ast \to \bar{t}\bar{t}$), in fact, also yield same-sign $W$'s from the top decays. We thus use our recast of the search in Ref.~\cite{Sirunyan:2018yun} to obtain a bound of  $M_{\pi_6} > 1.24$~TeV and an exclusion projection of $M_{\pi_6} > 1.62$~TeV for HL-LHC. These results are based on a LO simulation, with production cross section normalised at LO.
This result, shown by the red horizontal lines in Fig.~\ref{fig.pi6}, implies that the bounds from direct production of $\pi_6$ are stronger than those from the $X_{5/3}$ decays.

\subsection{Bounds for $X_{5/3}\rightarrow t\phi^{+}$}\label{sec.tphi+}

\begin{figure}[htbp]
	\begin{center}
		\includegraphics[width=0.45\textwidth]{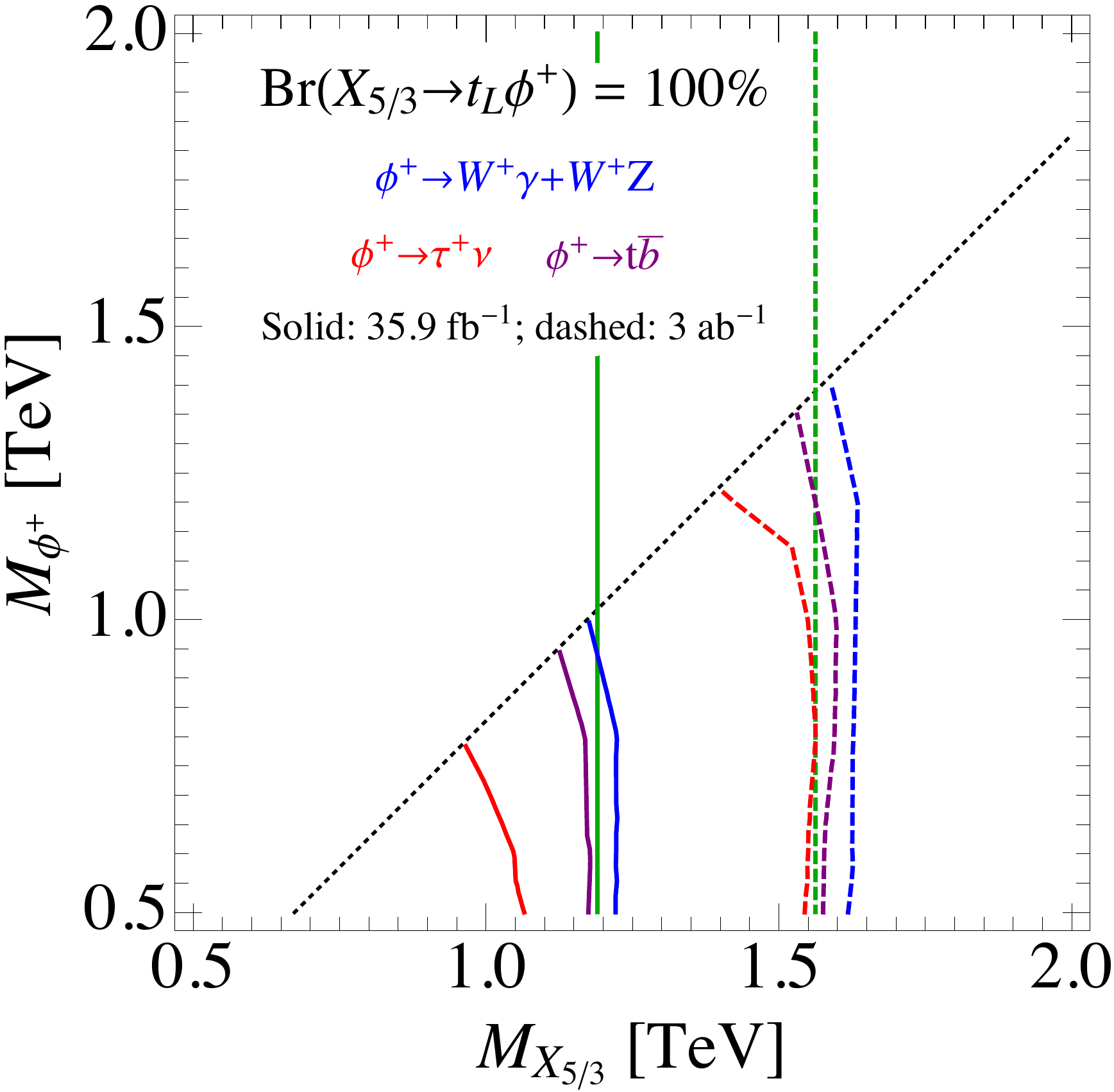}
		\caption{
			Bounds on the $M_{X_{5/3}}$--$M_{\phi^+}$ plane from the CMS SSL search of Ref.~\cite{Sirunyan:2018yun} (solid lines) and projections for the HL-LHC reach (dashed lines) for the various $\phi^+$ decay modes (blue-red-purple). The black dotted line indicates $M_{X_{5/3}}=M_{\phi^+}+m_t$, above which the decay $X_{5/3} \to t \phi^+$ becomes kinematically unaccessible. The bounds and projections for the standard ${\rm Br}(X_{5/3}\to tW^+)=1$ decay are shown for reference in green lines. The bounds become insensitive to $M_{\phi^+}$ for light masses above threshold.}
		\label{fig.phip}
	\end{center}
\end{figure}

Next, we discuss the bounds from SSL searches on $X_{5/3}$ pair production if $X_{5/3}$ decays through singly-charged scalars. This model is described in Section~\ref{sec.chmod}. The singly-charged scalar $\phi^+$ has several possible decay modes, where typically one dominates over the others:
\begin{itemize}
	\item[a)] $X_{5/3} \to t \phi^+ \to b W^+ (W^+ \gamma + W^+Z)$,
	\item[b)] $X_{5/3} \to t \phi^+ \to t t \bar{b} \to b b \bar{b} W^+ W^+$,
	\item[c)] $X_{5/3} \to t \phi^+ \to b W^+ \tau^+ \nu/\bar{\nu}$.
\end{itemize}
All final states yield similar SSL rates from the leptonic decays of the $W$, except for the leptonic tau decays in case c) that occur at a higher rate. As in the previous case, the main differences originate from the kinematics of the events: for instance, the leptons from the tau decay are expected to be softer because of the presence of an additional neutrino.
In the case a), the $\gamma/Z$ are predicted to be roughly $80\%/20\%$ due to the fact that $\Br(\phi^+\to W^+\gamma)/\Br(\phi^+\to W^+Z) \approx 1/\tan^{2} \theta_W\simeq 4$ when $M_{\phi^+}\gg m_{W,Z}$.

We simulated the new channels assuming 100\% branching ratio into each, and taking the chiral couplings predicted in the composite Higgs models.
Fig.~\ref{fig.phip} shows the resulting limits (solid) and the projected HL-LHC exclusion reaches (dashed). As it can be seen, the bounds are comparable to those obtained for the standard decay $X_{5/3}\to t W^+$. Constraints and projections for the decay $X_{5/3}\to t \phi^+ \to t \tau^+ \nu$ are slightly weaker, in particular for heavy $\phi^+$. This is because the leptons from $\tau$ decays are softer. Overall, as can be seen, the cuts applied in the SSL search are rather insensitive to the detailed kinematics. Thus, no large modifications of the bounds are expected even in case $X_{5/3}$ and $\bar{X}_{5/3}$ decay through different channels.

It is instructive to check kinematic distributions in order to determine whether and how the  different $X_{5/3}$ decays could be distinguished if an excess is found in future searches. The most striking feature is present in the $X_{{5/3}}\to t\phi^+\to tW^+\gamma$ decay channel, where the SSL is accompanied by a hard photon. We study this channel in more detail in Section~\ref{sec.SSLgam}.

\subsection{Bounds for $X_{5/3}\rightarrow b\phi^{++}$}\label{sec:phipp}

The model described in Section~\ref{sec.chmod} also contains a doubly charged scalar $\phi^{++}$, that allows for the $X_{5/3}\rightarrow b \phi^{++}$ decay. The doubly charged $\phi^{++}$ has several possible decays, with usually one of them dominating over the others:
\begin{itemize}
	\item[a)] $X_{5/3} \to b \phi^{++} \to b W^+ W^+$,
	\item[b)] $X_{5/3} \to b \phi^{++} \to b \tau^+ \tau^+$,
	\item[c)] $X_{5/3} \to b \phi^{++} \to b W^{+(\ast)} \phi^+$,
\end{itemize}
where the latter contains a virtual $W^+$ and several possible decays of the $\phi^+$. As before, all decay modes offer SSLs in the final state. There are two important features distinguishing this case form the previous ones: the $b\tau^+\tau^+$ decay offers higher leptonic rates compared to the $W$'s in the final state, without loosing too much hardness of the leptons; in the chain decay, one of the leptons needs to come from the virtual $W$, thus being typically very soft due to the small mass splitting between the two charged scalars. Furthermore, the $bW^+W^+$ final state tends to have harder SSL pairs and larger $H_T$, thus it more easily passes the selection cuts.

\begin{figure}[htbp]
	\begin{center}
		\includegraphics[width=0.45\textwidth]{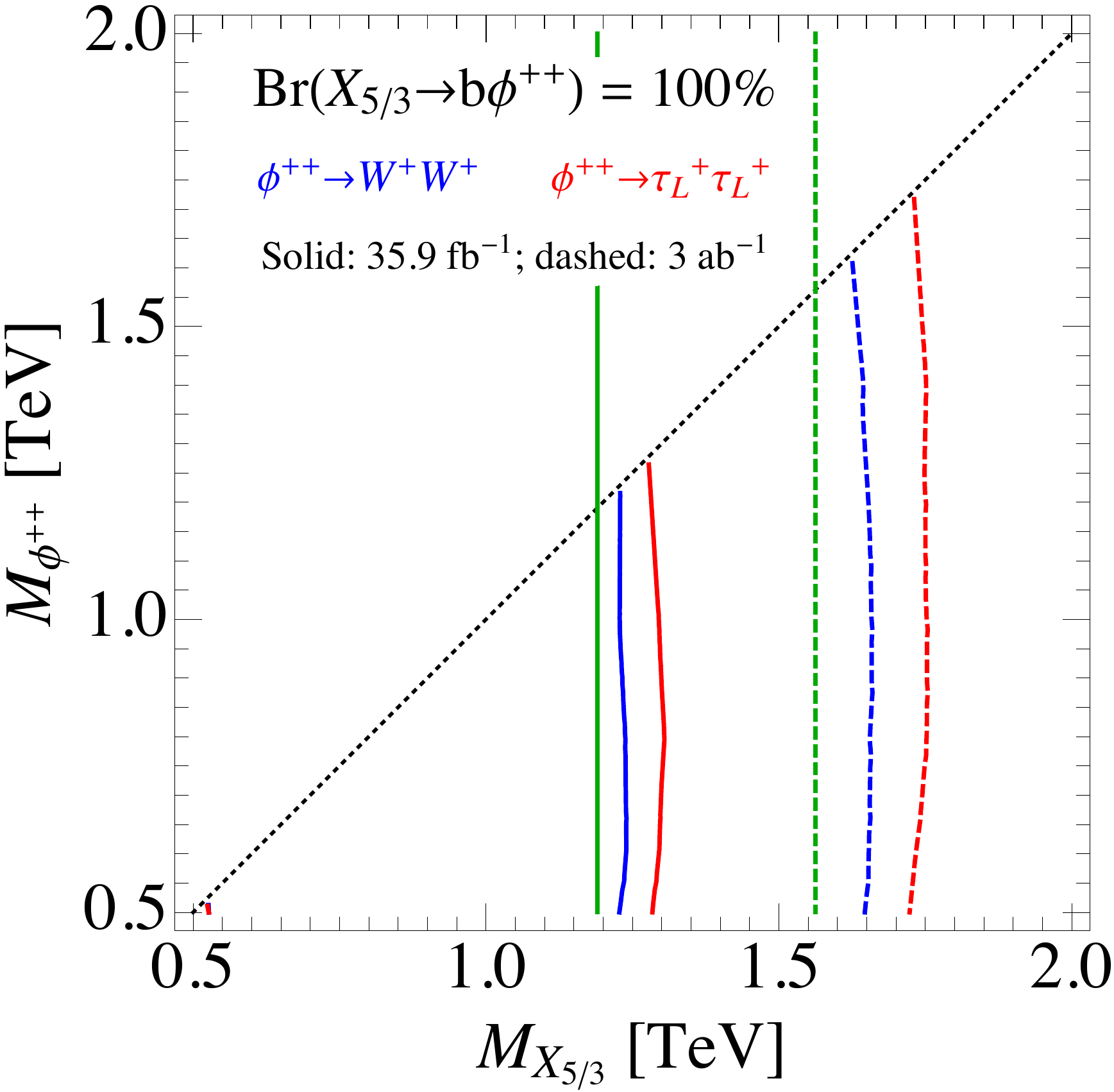} \qquad
		\includegraphics[width=0.45\textwidth]{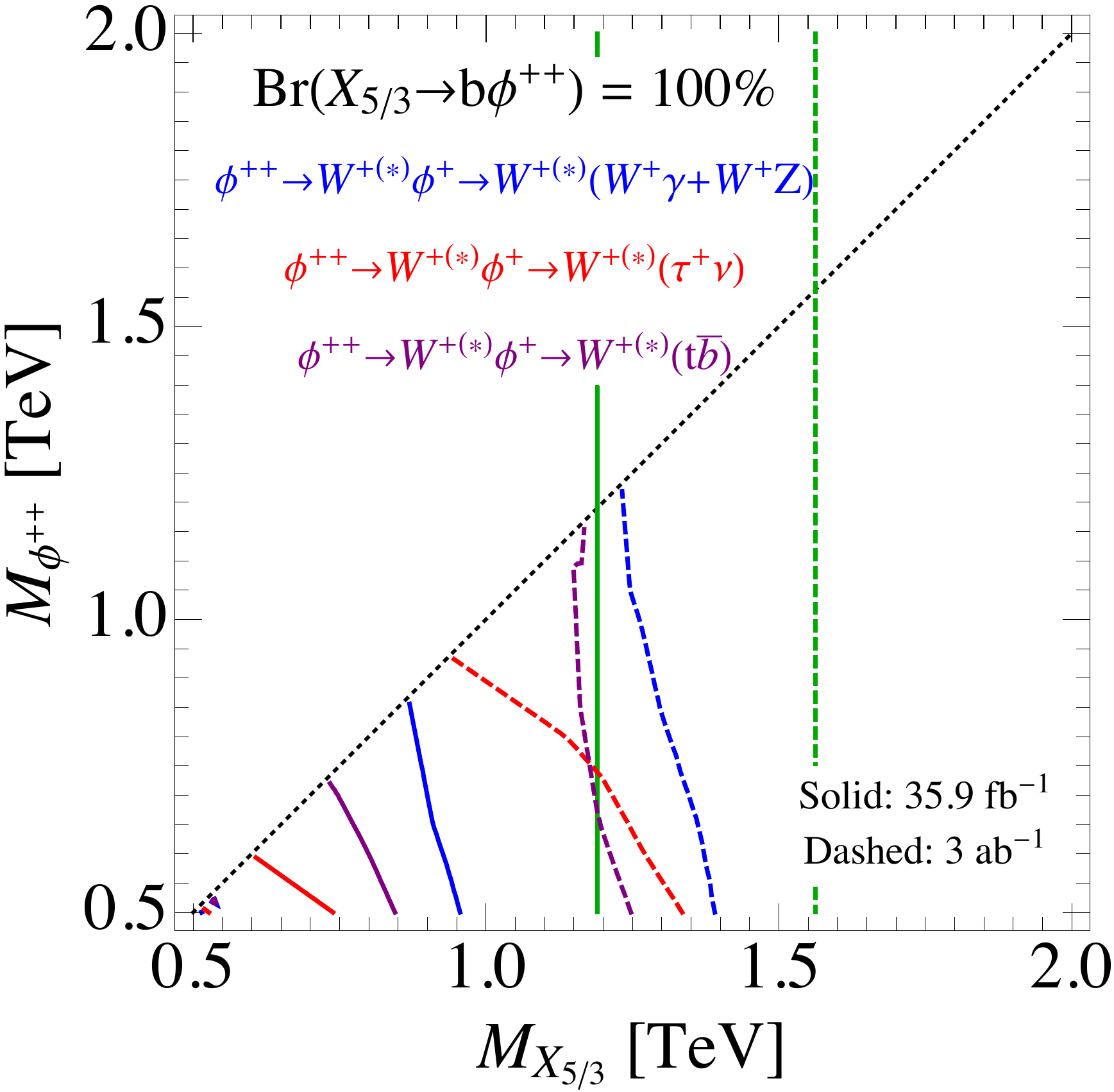}
		\caption{Bounds on the $M_{X_{5/3}}$--$M_{\phi^{++}}$ plane from the CMS SSL search of Ref.~\cite{Sirunyan:2018yun} (solid lines) and projections for the HL-LHC reach (dashed lines) for the various $\phi^{++}$ decay modes. In the left panel we show direct decays of $\phi^{++} \to W^+ W^+/\tau^+ \tau^+$, while in the right panel final states from the cascade decay $\phi^{++} \to \phi^+ W^{+*}$. The bounds and projections for the standard ${\rm Br}(X_{5/3}\to tW^+)=1$ decay are shown for reference in green lines.  The bounds become insensitive to $M_{\phi^{++}}$ for light masses above threshold.}
		\label{fig.phipp}
	\end{center}
\end{figure}	 

The results of our recast of the SSL search are shown in Fig.~\ref{fig.phipp}, where we present the direct decays of $\phi^{++}$ in the left panel and the chain decay via $\phi^+$ in the right one. As expected, the bounds for the direct decays are significantly stronger, with the highest gain in the $\tau^+ \tau^+$ channel. In the case of the chain decay, the bounds depend significantly on the mass split between the two charged scalars, which is expected to be small in the underlying models as they both come from the same custodial multiplet. Following Ref.~\cite{Agugliaro:2018vsu}, we expect the mass  difference to be a few tens of GeV, and in the simulation we fix $M_{\phi^{++}} - M_{\phi^+} = 30$~GeV. The results from Fig.~\ref{fig.phipp} thus show that the bounds on the masses are weaker than for the standard decay and for other decays of the charged scalars. As we already mentioned, this is due to the softness of the lepton coming from the virtual $W$ in $\phi^{++} \to\ell^+ \nu \phi^+$.
We should recall, however, that this final state will be covered by the SLj search that, as we mention, has similar expected reach as the SSL one we implemented.

%%%%%%%%%%%%%%%%%%%%%%%%%%%%%%%%%%%%%%%%%%%%%%%%%%%%%%
%%%%%%%%%%%%%%%%%%%%%%%%%%%%%%%%%%%%%%%%%%%%%%%%%%%%%%

\section{Opportunities: optimising specific channels}\label{sec.SSLgam}

As showed in the previous section, the SSL search for $X_{5/3}$ pair production applies very well to the motivated exotic $X_{5/3}$ decays, with comparable signal efficiency. Kinematic distributions of exotic $X_{5/3}$ decays, however, differ and can in principle be used to distinguish the signals if an excess is found, or optimise the cut-flow for specific cases. One signature which stands out is the $X_{5/3}\rightarrow tW^+\gamma$ decay through $\phi^+$ that yields an additional hard photon. \footnote{The same applies to the cascade decay $X_{5/3}\rightarrow b \phi^{++}\rightarrow bW^{+(*)}\phi^+\rightarrow b W^{+(*)} W^+\gamma$.} In this section we will provide a brief characterisation of the opportunities these kinematical features may offer for future searches.

\subsection{$X_{5/3}\rightarrow t\phi^+\rightarrow tW^+\gamma$: A hard photon final state}

For decays of the singly-charged scalar via the WZW coupling, there is a high probability (roughly 80\%) of having a hard photon in the final state.
Demanding a hard photon in addition to the SSL cut-flow strongly reduces the background while merely affecting the signal efficiency, thus it may be an efficient tool to further probe this decay mode. The photon in the final state can also be very effective to search for $\phi^+$ in direct Drell-Yan production,
as suggested for the Georgi-Machacek model in Ref.~\cite{Logan:2018wtm}. It should be noted that in the composite case this is relevant irrespective of the mass of the scalar, while in the elementary Georgi-Machacek model only for masses below the $WZ$ threshold.

Estimating the backgrounds for a SSL + hard photon search is, however, hard and beyond the scope of this work. The main reason is that we expect new sources of backgrounds deriving from instrumental misidentification of jets into photons, that can only be estimated reliably by data-driven methods. While we expect such backgrounds (and their systematic uncertainties) to be small, they may be potentially relevant for the HL-LHC study. Other SM backgrounds, however, can be easily estimated. In Ref. \cite{Sirunyan:2018yun}, the dominant backgrounds come from SM processes ($t\bar{t}+X$ and multi-boson final states) and same-sign non-prompt background (in which a non-prompt lepton, i.e. a lepton from a heavy-flavour decay, photon conversion, or a misidentified jet, passes the tight lepton identification requirements), while misidentified opposite sign prompt lepton background (in which one lepton charge is misidentified) is present but less important. Overall, about $45$ background events are expected (divided as $10.9\pm 1.9$, $11.2\pm 2.0$ and $23.2\pm 3.7$ events  in the $ee$, $\mu\mu$ and $e\mu$ channels) for a luminosity of $\mathcal{L}=35.9$~fb$^{-1}$.
Adding a radiated hard photon would bring down the rates by a factor $\alpha_{\rm em}$ times a loop factor, thus we expect a background suppression of $10^{-3} \div 10^{-4}$ roughly. This would leave a background free search at current luminosity, and very few background events at HL-LHC. As already mentioned, this very naive estimate suffers from the presence of additional instrumental backgrounds that can only be estimated using data-driven techniques.

\begin{figure}[htbp]
	\begin{center}
		\includegraphics[width=0.465\textwidth]{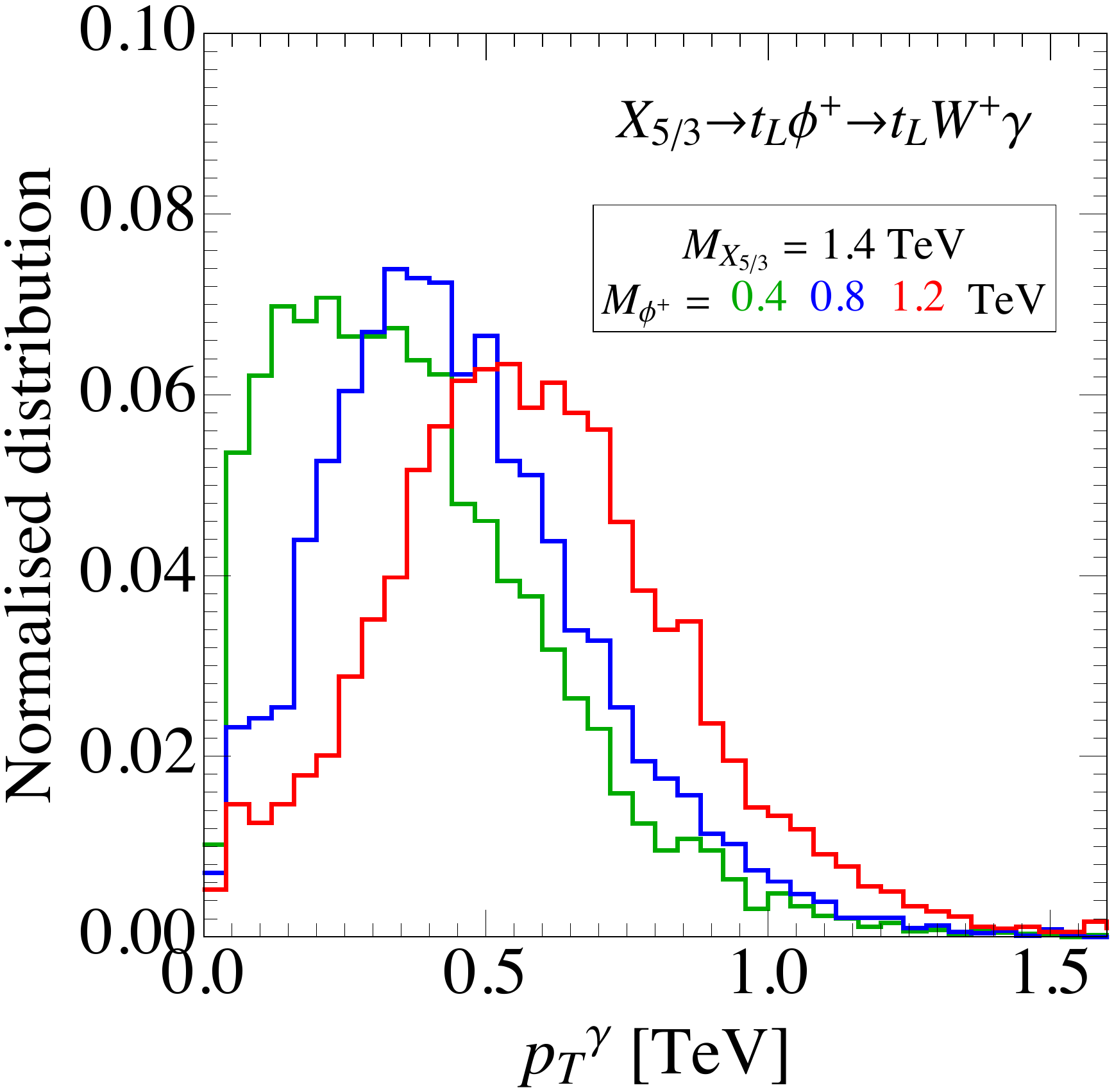}\qquad
		\includegraphics[width=0.45\textwidth]{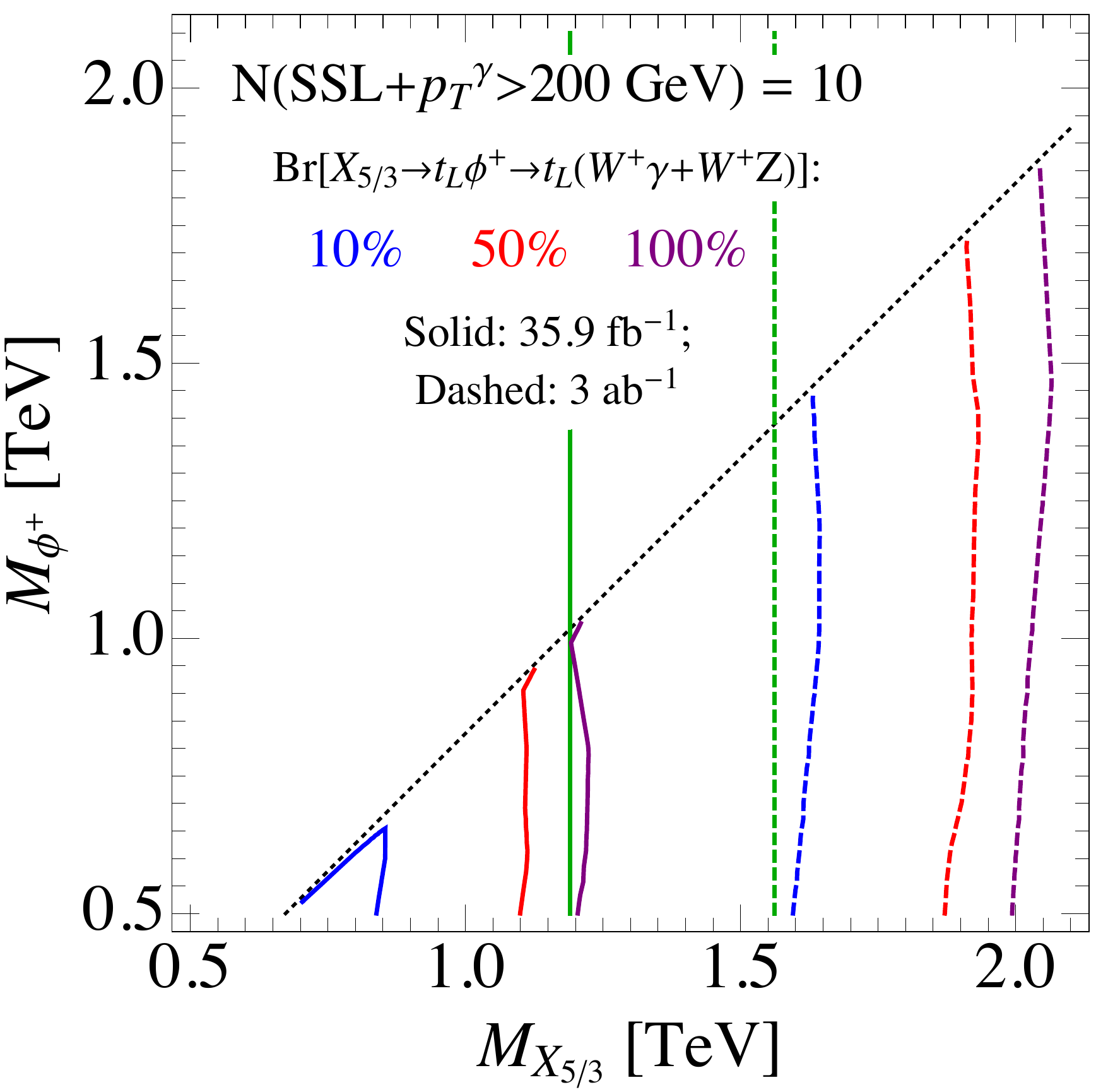}
		\caption{Left: Distribution of the transverse momentum of the leading photon in signal events which pass all SSL cuts, for $M_{X_{5/3}} = 1.4$~TeV and three reference masses of $M_{\phi^+}= 0.4, \, 0.8,$ and $1.2$~TeV.
			Right: Contours in the $M_{X_{5/3}} - M_{\phi^+}$ plane at which 10 signal events are expected with the cut-flow described in this subsection. Blue, red, and purple lines correspond to a $\Br(X_{5/3}\rightarrow t\phi^+)=10\%, \, 50\%, 100\%$. Solid lines assume 35.9 fb$^{-1}$ while dashed lines assume 3 ab$^{-1}$. The bounds and projections for the standard ${\rm Br}(X_{5/3}\to tW^+)=1$ decay are shown for reference in green lines.}
		\label{fig.photons}
	\end{center}
\end{figure}

For the reasons above, we do not attempt to estimate discovery/exclusion reaches, but we will only indicate the level of signal events that may be achieved. The search strategy is to use the same SSL cut-flow from Ref. \cite{Sirunyan:2018yun}, which has been outlined in detail in Section \ref{sec.recast}, and require an additional high-$p_T$ photon. As an illustration, in the left panel of Fig.~\ref{fig.photons} we show the $p^\gamma_T$ distributions of signal events passing all SSL cuts prior to the $p^\gamma_T$ cut, where we fixed $M_{X_{5/3}} = 1.4$~TeV and chose three reference masses of $M_{\phi^+}= 0.4, \, 0.8,$ and $1.2$~TeV. As expected, the photon spectrum becomes harder for heavier $\phi^+$ masses, however we can see that even light scalars tend to produce high-$p_T$ photons due to the large mass of the $X_{5/3}$ mother particle.
We thus decided to impose a cut
\begin{equation}
	p_T^\gamma > 200~\mbox{GeV}\,,
\end{equation}
as a reference value, which could be optimised in a more realistic search definition that included backgrounds.
With this choice, the majority of the signal events passes the cut while, as we argued before, the backgrounds should be greatly reduced. In the right panel of Fig.~\ref{fig.photons} we show, in the $M_{X_{5/3}}- M_{\phi^+}$ plane, contours at which 10 events pass the SSL + photon selection cuts for the luminosity of 35.9 fb$^{-1}$ from Ref.~\cite{Sirunyan:2018yun} (solid) and for the HL-LHC one (dashed). We probed 3 different values of the $\Br (X_{5/3} \to t \phi^+) = 10\%$, $50\%$ and $100\%$, with the remaining decays into the standard one $\Br(X_{5/3} \to t W^+) = 1 - \Br (X_{5/3} \to t \phi^+)$ (we recall that $\phi^+ \to W^+ \gamma (80\%) + W^+ Z (20\%)$). As a reference the green vertical lines show the current and projected reaches of the SSL search: while the two sets of curves cannot be directly compared as they correspond to very different quantities, the plot illustrates the fact that adding the requirement of a hard photon cut may improve the reach at HL-LHC, even for small branching ratios in the exotic $X_{5/3} \to t \phi^+$ channel. As we already stressed, while 10 events is a generous requirement at low luminosity, where zero background events are expected, only a dedicated experimental study can really establish the potential reach at HL-LHC.

As a final remark, we would recommend that any search sensitive to $X_{5/3}$ pair and single production, like Refs~\cite{Aaboud:2018uek,Aaboud:2018xpj,Sirunyan:2018yun}, could publish the $p_T$--distribution of the hardest photon as a first probe. The observation of an anomaly, c.f. Fig.~\ref{fig.photons}, would then be a hint that this exotic decay may be hidden in the background.

\subsection{Top-rich final states: jet and $b$-jet multiplicities}\label{sec.distributions}

\begin{figure}[htbp]
	\begin{center}
		\includegraphics[width=0.45\textwidth]{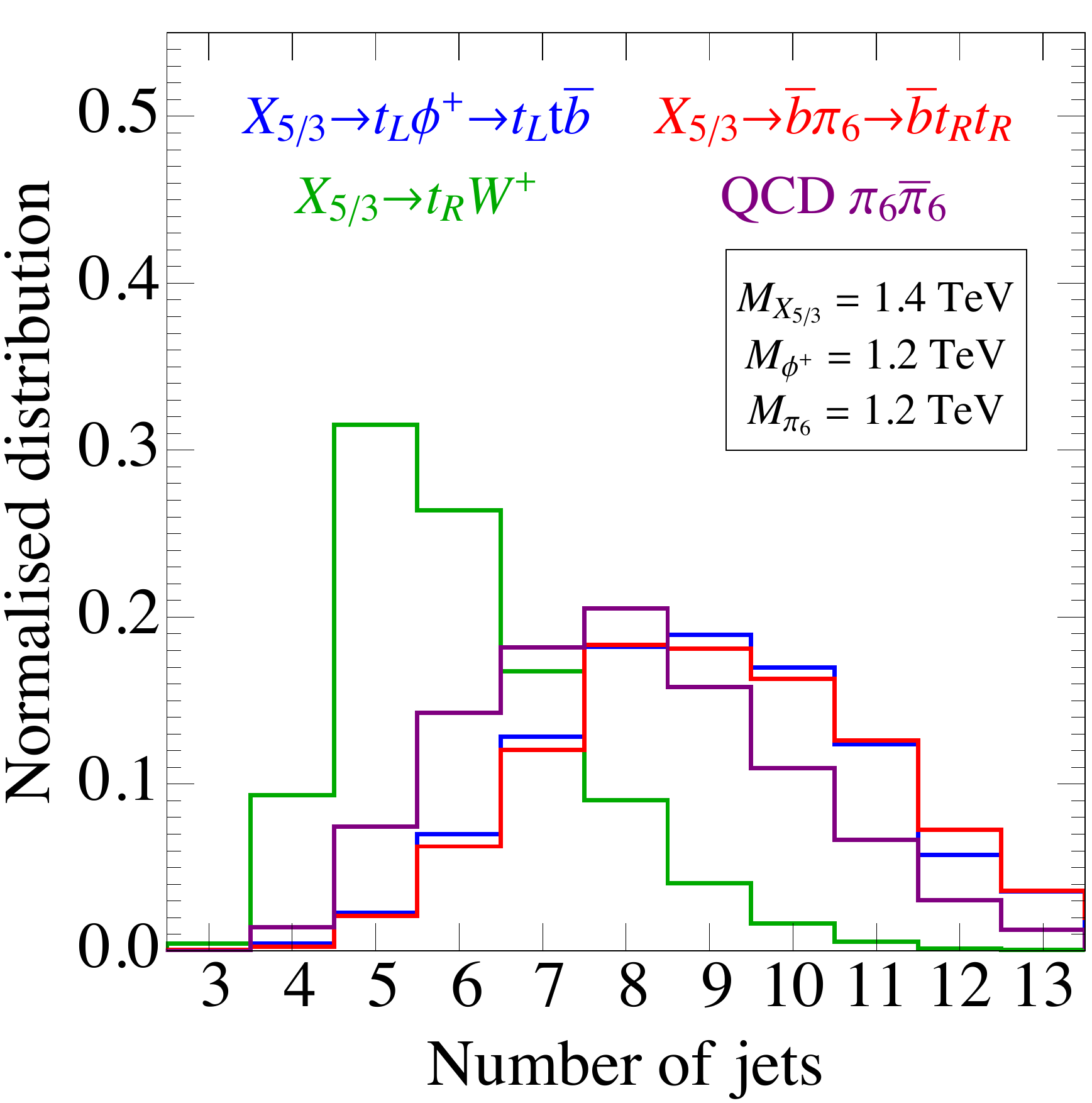}\qquad
		\includegraphics[width=0.45\textwidth]{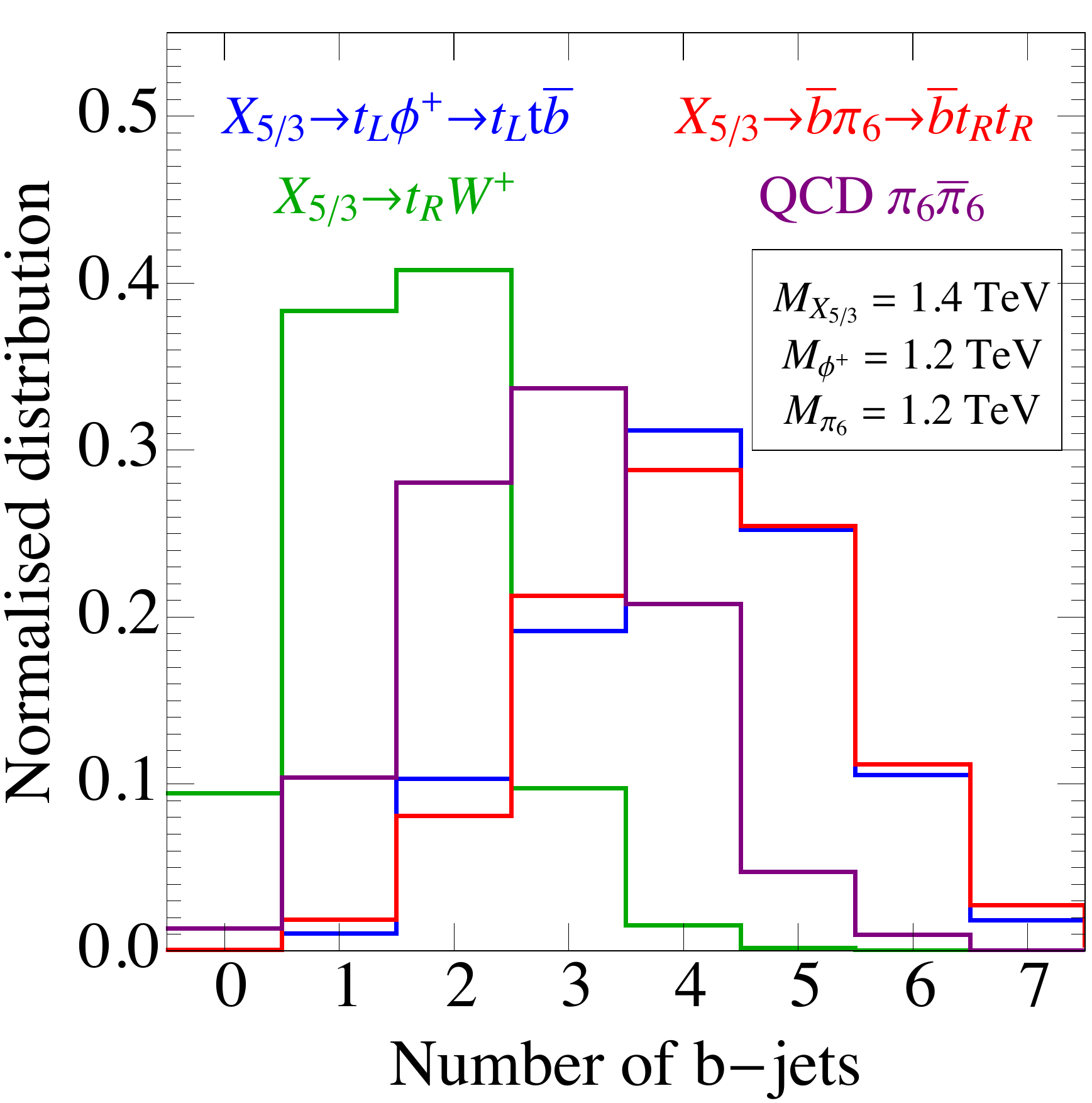}
		\caption{Number of jets (left) and $b$-jets (right) for various decay modes of the $X_{5/3}$ and for $\pi_6$ pair production.  The distributions are obtained after the SSL search cuts (see Section~\ref{sec.recast}).}
		\label{fig.Njet}
	\end{center}
\end{figure}

Several channels of the exotic $X_{5/3}$ decays yield more tops in the final state than the standard decay. It is well known that such final states are efficiently detected by SSL searches~\cite{Deandrea:2014raa}, as our results in the previous section also confirm. Upon decays of the tops, the final state results very rich in $b$-jets and has enhanced hadronic activity. The same holds true for the  $\pi_6$ QCD pair production, for which we applied the SSL search in Section~\ref{sec:pi6}. Thus, it may be useful to explore the $b$-jet and light-jet multiplicities to distinguish the exotic channels from the standard one.
To this end, we show these distributions in Fig.~\ref{fig.Njet} for 
\begin{itemize}
	\item $X_{5/3}$ pair production and the ``standard'' decay $X_{5/3}\rightarrow t W^+$,
	\item $X_{5/3}$ pair production and the charged cascade decay $X_{5/3}\rightarrow t \phi^+\rightarrow tt\bar{b}$,
	\item $X_{5/3}$ pair production and the ``coloured'' decay  $X_{5/3}\rightarrow \bar{b} \pi_6 \rightarrow \bar{b} t t$, and
	\item $\pi_6$ pair production with $\pi_6\rightarrow t t$. 
\end{itemize}
For the $b$-tagging, following Ref.~\cite{Aaboud:2018xpj}, we choose an efficiency of $77\%$, with mis-tag rates of  $1/134$ ($1/6$) for light jets ($c$-jets), implemented in the analysis by modifying the {\tt Delphes} card.
The distributions show events which pass all SSL cuts as outlined in Section~\ref{sec.recast}. We fixed the masses of the new  states to $M_{X_{5/3}}=1.4$~TeV, $M_{\phi^+}=1.2$~TeV, and $M_{\pi_6}=1.2$~TeV, however the distributions only weakly depend on the masses. The jets and $b$-jets result from (cascade) decays of fairly heavy particles such that they are very likely to overcome basic energy or $p_T$ cuts. A mild dependence is introduced as hadronically decaying tops or $W$ bosons are increasingly boosted for higher masses, leading to an increasing rate of merged jets or $b$-jets.

As expected, the plots show that the exotic signals feature more $b$-tagged jets and hadronic activity than the standard signal. Besides a discriminator, this feature could be used in the HL-LHC data to increase the level of background rejection against these new signals.

\subsection{Modified kinematics of the SSL pair}\label{sec.distributions}

\begin{figure}[htbp]
	\begin{center}
		\includegraphics[width=0.45\textwidth]{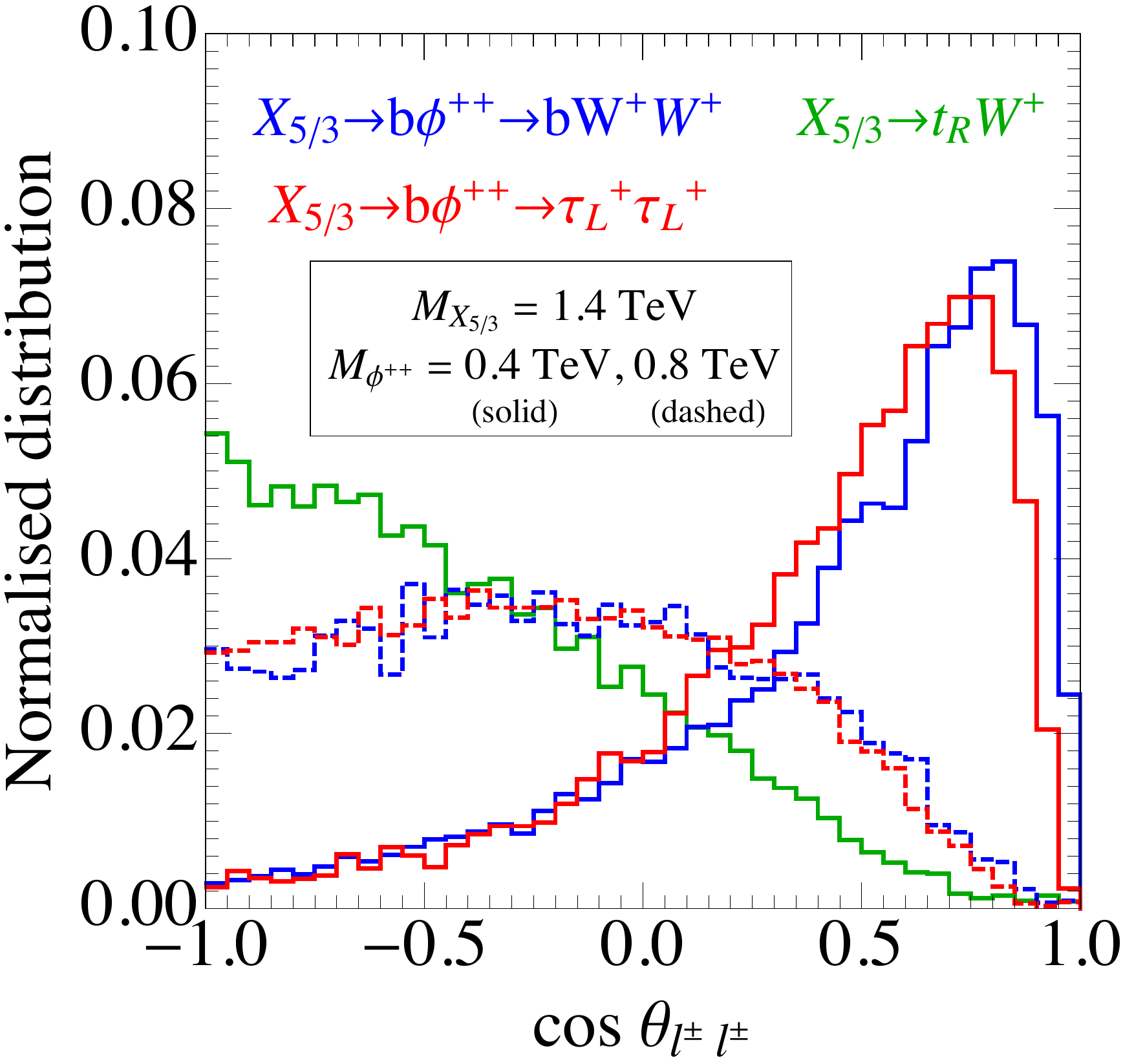}\qquad
		\includegraphics[width=0.45\textwidth]{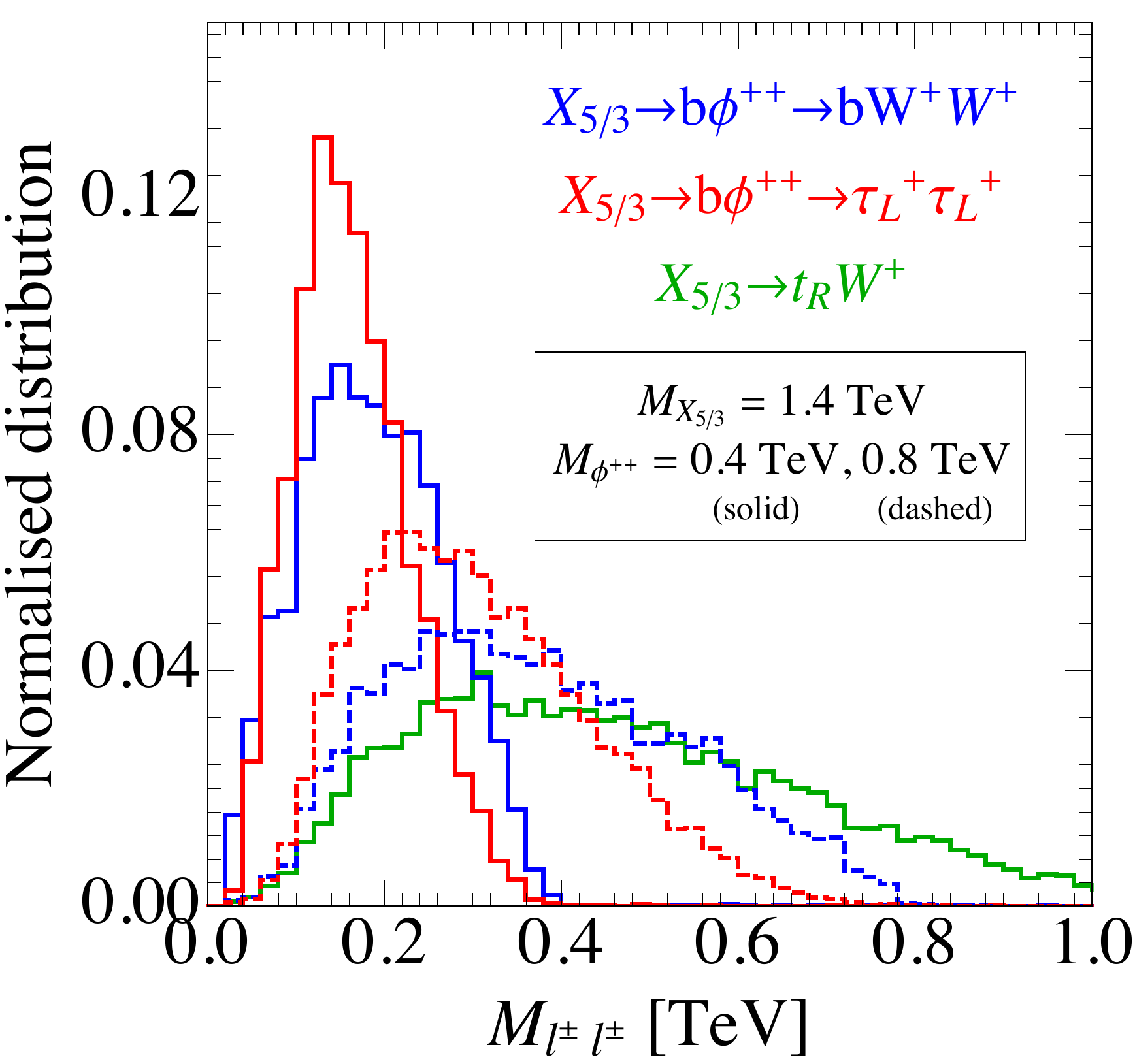}
		
		\caption{Distribution of the relative azimuthal angle between the SSL pair (left) and invariant mass (right) for various $X_{5/3}$ decays, assuming 100\% branching ratio. All events pass the SSL cuts detailed in Section~\ref{sec.recast}.}
		\label{fig.phippdist}
	\end{center}
\end{figure}

The exotic $X_{5/3}$ decays alter the kinematics of the final state, which in particular affects the SSL kinematic distributions. 
In this section, we focus on two channels that have the most striking effects: $X_{5/3} \to b \phi^{++} \to b W^+ W^+$ and $X_{5/3} \to b \phi^{++} \to b \tau^+ \tau^+$. We show the distribution, for exclusive decays, in Fig.~\ref{fig.phippdist}, compared to the standard decay $X_{5/3} \to t W^+$.

As we are considering QCD pair production, the majority of the top partners will be produced at rest and do not have a significant momentum. For the standard decay, the top and $W^+$ will therefore be produced back-to-back and with a sizeable boost. This implies that the SSL pair, where one lepton comes from the top and the other from the $W$, also tends to be back-to-back, as shown in the left panel of Fig.~\ref{fig.phippdist}. On the other hand, for the decays via the doubly-charged scalar $\phi^{++}$, the SSL pair comes from a single boosted resonance, and thus tends to be more collinear. This effect is more pronounced for light scalars: in the figure, we show two sets of distributions for fixed $M_{X_{5/3}} = 1.4$~TeV and two choices of $M_{\phi^{++}} = 800$~GeV and $M_{\phi^{++}} = 400$~GeV. The plot clearly shows that increasing the scalar mass makes the signal more similar to the $tW^+$ case, while a net distinction is exhibited for small masses. 

We also see a marked effect in the invariant mass distribution of the SSL pair. For light $\phi^{++}$, the scalar will receive a momentum of roughly $p_{\phi^{++}} \approx (M_{X_{5/3}} - M_{\phi^{++}})/2$. In the case of $\phi^{++} \to \tau^+ \tau^+ \to \ell^+ \ell^+ + 4 \nu$'s, we can roughly expect that this momentum will be equally subdivided between the 6 particles in the final state, so that the expected invariant mass should be around $M_{\ell^\pm \ell^\pm} \approx (M_{X_{5/3}} - M_{\phi^{++}})/6$, as it can be seen in Fig.~\ref{fig.phippdist}, right panel, for $M_{\phi^{++}} = 400$~GeV. A larger invariant mass can be expected for the $\phi^{++} \to W^+ W^+$ channel due to the presence of fewer neutrinos in the final state. For the standard decay, due to the large angular separation, we expect and see a broader distribution, with larger invariant masses being populated.

These results show that the angular separation and invariant mass can be good discriminants in the case of decays via the doubly charged scalar, in particular for light masses compared to the top partner one.

%%%%%%%%%%%%%%%%%%%%%%%%%%%%%%%%%%%%%%%%%%%%%%%%%%%%%%
%%%%%%%%%%%%%%%%%%%%%%%%%%%%%%%%%%%%%%%%%%%%%%%%%%%%%%

\section{Conclusions}\label{sec.conclusions}

In realistic models of a composite Higgs boson, the top partners have more available decays than the standard ones in a pair of SM particles, which have been considered so far. We have focused our attention on the decays of the custodial charge $5/3$ partner $X_{5/3}$, which has only one possible decay into a SM final state, $X_{5/3} \to t W^+$. 
The new channels always involve at least a pair of same-sign tops or $W$ bosons, which lead to the SSL signature already searched for in the standard decay mode for pair produced $X_{5/3}$.

After recasting the search in Ref. \cite{Sirunyan:2018yun}, we showed that all the final states have similar efficiencies, thus leading to very similar bounds on the mass of the $X_{5/3}$, irrespective of the precise branching ratios. The only exception is due to chain decays through a doubly-charged and a singly-charged scalar, with the two charged scalars being close in mass, which however will be covered by the single-lepton searches.

While the new channels are already constrained, their final states, richer than the standard one, also offer opportunities for improvements of the current strategies. Such improvements may be crucial at the HL-LHC phase. In particular, we identified one decay via the singly-charged scalar which contains high-$p_T$ photons, $X_{5/3} \to t \phi^+ \to t W^+ \gamma$. Adding a high-$p_T$ photon requirement to the standard SSL search may be able to consistently increase the reach at the LHC, even though a data-driven background estimation is needed for a precise determination.
The new channels also offer final states rich in tops, which therefore feature many $b$-tagged jets and increased hadronic activity, while some channels feature peculiar angular distribution of the SSL pair that may allow to distinguish them from the standard decays.

In conclusion, we showed that the new exotic decay channels, which are the norm in realistic models, while being already efficiently covered, offer new opportunities for improvement for HL-LHC searches of the custodial $X_{5/3}$ top partner.

%%%%%%%%%%%%%%%%%%%%%%%%%%%%%%%%%%%%%%%%%%%%%%%%%%%%%%
%%%%%%%%%%%%%%%%%%%%%%%%%%%%%%%%%%%%%%%%%%%%%%%%%%%%%%

\acknowledgments

We thank Rui Zhang and Mengchao Zhang for useful discussions, and Yongcheng Wu  for communication on Ref.~\cite{Han:2018hcu}. KPX is supported by Grant Korea NRF 2015R1A4A1042542 and NRF 2017R1D1A1B03030820. TF's work is supported by IBS under the project code IBS-R018-D1.
GC acknowledges partial support from the France-Korea Particle Physics Lab (FKPPL) and the Labex-LIO (Lyon Institute of Origins) 
under grant ANR-10-LABX-66 (Agence Nationale pour la Recherche), and FRAMA (FR3127, F\'ed\'eration de Recherche ``Andr\'e Marie Amp\`ere'').

%%%%%%%%%%%%%%%%%%%%%%%%%%%%%%%%%%%%%%%%%%%%%%%%%%%%%%
%%%%%%%%%%%%%%%%%%%%%%%%%%%%%%%%%%%%%%%%%%%%%%%%%%%%%%

\appendix

\section{Bounds on an exotic decay of a top partner with charge 2/3}\label{app:Tta}

\subsection{$T_{2/3}\rightarrow t a \rightarrow tt\bar{t}$}
In the main article we focussed on exotic decays of charge 5/3 partners. However, as is well known, SSL searches yield strong bounds on more generic multi-top final states~\cite{Deandrea:2014raa}.  In particular, this statement holds for charge $2/3$ top partners, $T_{2/3}$, that can decay to 3 tops, $tt\bar{t}$, via a bosonic mediator~\cite{Bizot:2018tds}. This case has already been considered via vectors~\cite{Deandrea:2014raa} and scalars~\cite{Han:2018hcu}, but with older data. Thus, as a byproduct of our $X_{5/3}$ study, we will apply our recast search to this final state.

We will consider the following effective Lagrangian~\cite{Bizot:2018tds} (implemented as described in Section~\ref{sec.recast})
\bea
\mathcal{L}_{T_{2/3},a}^{t\bar{t}} &=&\bar{T}_{2/3} \left(i \slashed{D} - M_{T_{2/3}}\right)T_{2/3} 
+ \left(\kappa^T_{a,L}  \, a \bar{T}_{2/3}P_L t + \mbox{h.c.} \right) \nonumber\\
&&+ \frac{1}{2}\partial_\mu a \partial^\mu a - \frac{1}{2}M_a^2 a^2 - i C_t a \bar{t} \gamma^5 t\,,
\eea
where $a$ is a new pseudo-scalar that decays 100\% into $t\bar{t}$, thus leading to the chain decay $T_{2/3} \to t a \to t t \bar{t}$.
While our Lagrangian, motivated by realistic composite Higgs models, differs from the ones considered in Refs.~\cite{Deandrea:2014raa,Han:2018hcu}, we expect the bounds not to depend much on the spin and CP-properties of the mediator.

\begin{figure}[htbp]
	\begin{center}
		\includegraphics[width=0.45\textwidth]{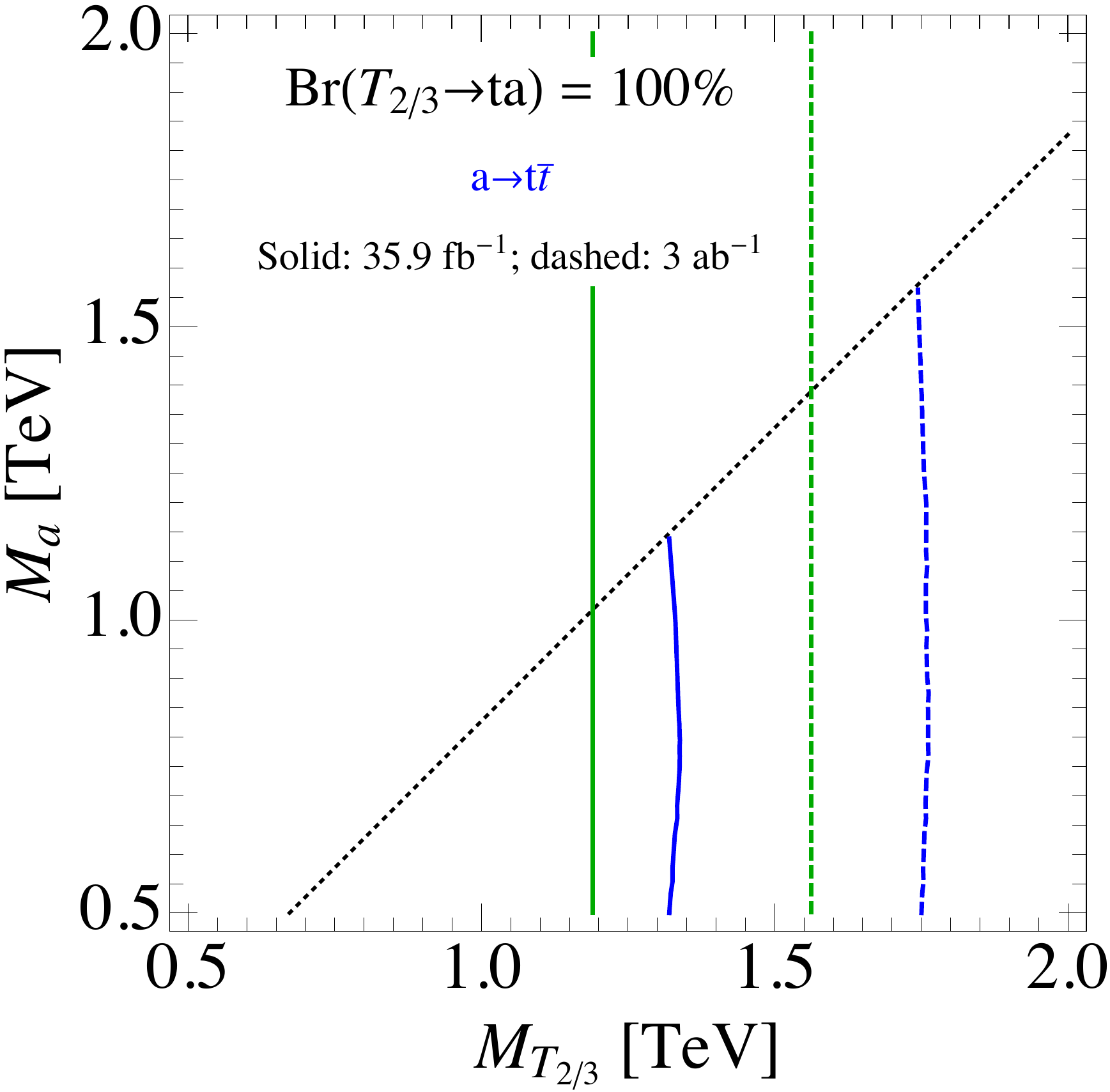} \hspace{0.4cm}
		\includegraphics[width=0.45\textwidth]{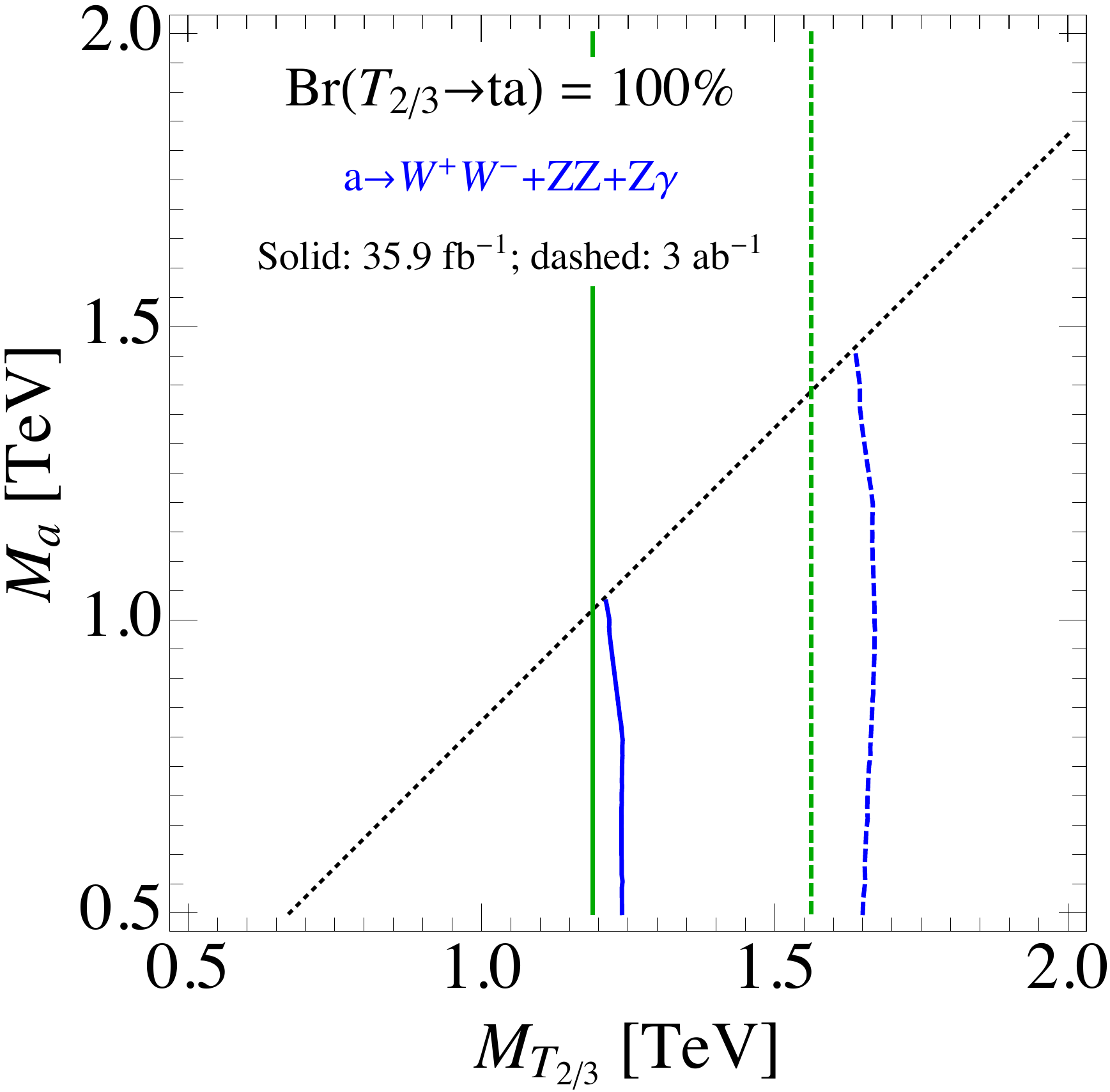}
		\caption{Bounds on the $M_{T_{2/3}}$--$M_{a}$ plane from the CMS SSL search of Ref.~\cite{Sirunyan:2018yun} (solid lines) and projections for the HL-LHC reach (dashed lines). On the left we show the exotic decay $T_{2/3}\to t a \to t t\bar{t}$, on the right the exotic decay $T_{2/3}\to t a \to t W^+W^-$ assuming the photo-phobic benchmark model with $\kappa_B = -\kappa_W$. The bounds and projections for $X_{5/3}$ pair-production with ${\rm Br}(X_{5/3}\to tW^+)=1$ are shown for reference in green lines. The bounds become insensitive to $M_a$ for light masses above threshold.}
		\label{fig.T23}
	\end{center}
\end{figure}	 

The bounds from our recast of the SSL search of Ref.~\cite{Sirunyan:2018yun} are shown in the left panel of Fig.~\ref{fig.T23}, where the solid blue line corresponds to the current luminosity and the dashed one is the HL-LHC projection. For reference, we also show the bounds on $X_{5/3}$ pair-production with ${\rm Br}(X_{5/3}\to tW^+)=1$. The bounds and projected bounds on $T_{2/3} \rightarrow tt\bar{t}$ are stronger than those of $X_{5/3}\to tW^+$ because the branching fraction for a $3t 3\bar{t}$ to SSL is larger. Our recast, yielding bounds $M_{T_{2/3}} \lesssim 1.3$~TeV are stronger than the ones in Ref.~\cite{Han:2018hcu} as the authors only consider Run-II LHC searches at lower luminosity~\cite{TheATLAScollaboration:2016gxs,Aad:2016tuk}. Nevertheless, the projections for HL-LHC are substantially weaker than those of the dedicated search suggested in Ref.~\cite{Han:2018hcu}, which is tailored to the 6 top final state.

\subsection{$T_{2/3}\rightarrow t a \rightarrow tW^+W^-$}

SSL searches also yield bounds on the decay $T_{2/3}\rightarrow t a \rightarrow tW^+W^-$, which can occur in models in which $a$ decays via electroweak interactions. 

We will consider the following effective Lagrangian~\cite{Bizot:2018tds} (implemented as described in Section~\ref{sec.recast})
\bea
\mathcal{L}_{T_{2/3},a}^{\rm{ew}} &=&\bar{T}_{2/3} \left(i \slashed{D} - M_{T_{2/3}}\right)T_{2/3} 
+ \left(\kappa^T_{a,L}  \, a \bar{T}_{2/3}P_L t  + \mbox{h.c.} \right) \nonumber\\
&&+ \frac{1}{2}\partial_\mu a \partial^\mu a - \frac{1}{2}M_a^2 a^2 + \frac{g^2\kappa^a_W}{8 \pi^2 f_a} a W^+_{\mu\nu}\tilde{W}^{-,\mu,\nu} \\
&&+ \frac{e^2\kappa^a_\gamma}{16 \pi^2 f_a} a A_{\mu\nu}\tilde{A}^{\mu,\nu} + \frac{g^2 c^2_W \kappa^a_Z}{16 \pi^2 f_a} a Z_{\mu\nu}\tilde{Z}^{\mu,\nu} + \frac{egc_W \kappa^a_{Z\gamma}}{8 \pi^2 f_a} a A_{\mu\nu}\tilde{Z}^{\mu,\nu}, \nonumber
\eea
where $f_a$ denotes the pseudo-scalar decay constant and $c_W$ is the Weinberg angle. In underlying models where $a$ is realized as a pNGB bound state of electroweakly charged fermions, the couplings to gauge bosons can be determined in terms of two parameters: one coupling, $\kappa^a_W$, to the
$SU(2)$ bosons and one,  $\kappa^a_B$ to  hypercharge, and couplings are related by
\begin{equation}
\kappa^a_\gamma = \kappa^a_W + \kappa^a_B \, , \,\,\, \kappa^a_Z = \kappa^a_W + \kappa^a_B t_W^4\, , \, \kappa^a_{Z\gamma} = \kappa^a_W - \kappa^a_B t_W^2,
\end{equation}
while the coupling constant to $W^+W^-$ is $\kappa^a_W$. Thus, within underlying models, the decay $a\rightarrow W^+W^-$  implies the presence of decays into neutral gauge bosons. In the following we use as a benchmark $\kappa_B = -\kappa_W$. This coupling structure is for example realized for a pNGB in underlying composite Higgs models with $SU(4)/Sp(4)$ breaking in the electroweak sector. In this benchmark point, the coupling of $a$ to $\gamma\gamma$ vanishes, while the branching fractions to $WW:ZZ:Z\gamma$ are $0.65\, : \, 0.16 \, : \, 0.19$ (up to phase-space factors, i.e. in the limit $M_a \gg 2 m_Z$).

The bounds from our recast of the SSL search of Ref.~\cite{Sirunyan:2018yun} are shown in the right panel of Fig.~\ref{fig.T23}, where the solid blue line corresponds to the current luminosity and the dashed one is the HL-LHC projection. As compared to the standard decay $X_{5/3}\to tW^+$ with ${\rm Br}(X_{5/3}\to tW^+)=1$, the bounds are slightly enhanced. Although the ${\rm Br}(T_{2/3}\to tW^+W^-) < 1$ in the benchmark scenario, the final state contains one more $W$, which enhances the probability to obtain same-sign leptons.
The decays $T_{2/3}\rightarrow t (ZZ / Z\gamma / \gamma\gamma)$ provide additional promising exotic VLQ decay signatures (see Ref.~\cite{Benbrik:2019zdp}).

%\begin{figure}[htbp]
%	\begin{center}
%		\includegraphics[width=0.45\textwidth]{2t4W.pdf}
%		\caption{Bounds on the $M_{T_{2/3}}$--$M_{a}$ plane from the CMS SSL search of Ref.~\cite{Sirunyan:2018yun} (solid lines) and projections for the HL-LHC reach (dashed lines) for the exotic decay $T_{2/3}\to t a \to t W^+W^-$, assuming the photo-phobic benchmark model with $\kappa_B = -\kappa_W$ (blue). The bounds and projections for $X_{5/3}$ pair-production with ${\rm Br}(X_{5/3}\to tW^+)=1$ are shown for reference in green lines.}
%		\label{fig.T23W}
%	\end{center}
%\end{figure}	 

%%%%%%%%%%%%%%%%%%%%%%%%%%%%%%%%%%%%%%%%%%%%%%%%%%%%%%
%%%%%%%%%%%%%%%%%%%%%%%%%%%%%%%%%%%%%%%%%%%%%%%%%%%%%%
\newpage

\bibliography{VLQeX53}
\bibliographystyle{JHEP-2-2.bst}

\end{document}